\documentclass[10pt,preprint]{aastex}
\usepackage{color}

\newcommand{\etal}{et~al.}

\newcommand{\teff}{$T_{\rm eff}$}

\newcommand{\vsini}{$v \sin i$}
\newcommand{\mum}{$\mu$m}

\begin{document}

\title{A Catalog of Point Sources Towards NGC 1333}

\author{L.~M.~Rebull\altaffilmark{1}}

\altaffiltext{1}{Infrared Science Archive (IRSA) and Spitzer Science
Center (SSC), Infrared Processing and Analysis Center (IPAC), 1200 E.\
California Blvd., California Institute of Technology, Pasadena, CA
91125; rebull@ipac.caltech.edu}

\begin{abstract}  

I present a catalog of point source objects towards NGC 1333,
resolving a wide variety of confusion about source names (and
occasionally positions) in the literature. I incorporate data from
optical to radio wavelengths, but focus most of the effort on being
complete and accurate from $J$ (1.25 \mum) to 24 \mum. The catalog
encompasses  $52\arcdeg<$RA$<52.5\arcdeg$ and
$31\arcdeg<$Dec$<31.6\arcdeg$.  Cross-identifications include those
from more than 25 papers and catalogs from 1994-2014, primarily those
in wide use as origins of nomenclature. Gaps in our knowledge are
identified, with the most important being a lack of spectroscopy for
spectral types or even confirmation of youth and/or cluster
membership. I fit a slope to the spectral energy distribution (SED)
between 2 and 24 \mum\ for the members (and candidate members) to obtain
an SED classification, and compare the resulting classes to those for
the same sources in the literature, and for an SED fit between 2 and 8
\mum. While there are certainly differences, for the majority of the
sources, there is good agreement. \end{abstract}


\section{Introduction}
\label{sec:intro}

NGC~1333 is one of the youngest and most well-studied star forming
regions, in part because it is located at only $\sim$235 pc (Hirota
\etal\ 2008, 2011).  Its stars are thought to have an average age of
1-2 Myr (e.g., Bally \etal\ 2008), but it also contains several
Class 0 objects, objects in the earliest stages of star formation
(see, e.g., Sadavoy \etal\ 2014 or Sandell \& Knee 2001). 

Because this region is very young and relatively nearby, it has been a
subject of intensive study for decades. However, there has not yet
been a published, comprehensive merging of all of the large catalogs
in this region. A summary including each of the prior investigations
of NGC 1333 is beyond the scope of this paper; see Walawender \etal\
(2008) for a recent review. In this paper, my primary goal is to
merge all of the available data from the (relatively) large-field
surveys in this region, and assemble one master catalog with all of
the names from the various surveys reconciled. We are studying NGC
1333 as part of the Young Stellar Object VARiability (YSOVAR) project
(Rebull \etal\ 2014, hereafter R14); this catalog was assembled
originally as part of that effort.  Because my original goal in taking
on this task was to focus on the sources for which we have light
curves in YSOVAR, some of the data reduction relevant to this paper is
described in the YSOVAR overview paper (R14).  Some of the detailed
investigation described in the present paper is focused on the region
mapped by YSOVAR; the region mapped by YSOVAR is the heart of the
cluster. The NGC 1333 YSOVAR data are discussed in detail in Rebull
\etal\ (2015, in preparation).

The catalog is somewhat artificially limited to being within
$52\arcdeg<$RA$<52.5\arcdeg$ and $31\arcdeg<$Dec$<31.6\arcdeg$. Data
are available over a larger region for some surveys, but this region
should include most of the objects actually belonging to NGC 1333, and
it entirely includes the region monitored for YSOVAR, as per the
original goals in assembling this catalog. 

The primary reason I compiled data from several different sources was
in order to assemble spectral energy distributions (SEDs). While, of
course, many of the targets vary significantly with time, single-epoch
archival data can help define the SED such that, in some cases, the
assembled SED can reveal the underlying nature of the source, or at
least help narrow the possibilities for the nature of the source.  

In this paper, I first review the large surveys that I included in our
catalog (Sec.~\ref{sec:largedatasets}), and then list the papers from
which I drew data, and explain why I associated or dissociated (or
removed) sources across wavelengths (Sec.~\ref{sec:litdata}). I
describe obvious gaps in the literature (Sec.~\ref{sec:future}). I
place the YSO candidates into SED classes based on the slope between 2
and 24 \mum, and compare them to other schemes from the literature
(Sec.~\ref{sec:sed}). 

\begin{deluxetable}{cccp{8cm}}
\tabletypesize{\scriptsize}
\rotate
\tablecaption{Overview of Studies and Data Included\label{tab:allstudies}}
\tablewidth{0pt}
\tablehead{\colhead{Dataset} & \colhead{Year published (or obtained)}
&\colhead{Band(s)} & \colhead{notes}}
\startdata
ASR & 1994 &  $JHK$ & many coordinate issues; see text\\
LAL & 1996 &  $JHK$ & many coordinate issues; see text\\
Preibisch & 1997 & X-ray (ROSAT) & two coordinate issues; source 2 may
be spurious\\
VLA & 1999 & 3.6, 6 cm & hard to find short-wavelength counterparts\\
2MASS & (2000) & $JHK_s$ & data obtained in 2000 as part of 2MASS
all-sky survey and subsequently the 2MASS 6$\times$ survey\\
Getman (et al.)& 2002 & X-ray (Chandra/ACIS) & two inconsistencies\\
Preibisch & 2003 & X-ray (XMM) & most sources match Chandra sources\\
Rebull (et al.)& 2003 & MIR & primary information included from this work is source
multiplicities\\
Aspin & 2003 & NIR & spectroscopy; spectral types included \\
MBO & 2004 & $JHK$ & duplicate MBO number \\
c2d & 2006, 2007 & Spitzer: 3.6 to 70 \mum\ & included limits and 70 \mum\ sources\\
Hatchell (et al.)& 2007 & 450, 850 \mum\ & cross ids retained \\
Greissl (et al.) & 2007 & $JH$ & NICMOS brown dwarf search\\
Gutermuth (et al.) & 2008 & Spitzer: 3.6 to 24 \mum & only YSO candidates reported \\
Oasa  (et al.)& 2008 & $JHK_s$ & brown dwarf search \\
Scholz (et al.) & 2009, 2012ab & $i^{\prime}z^{\prime}JK$ & SONYC; brown dwarf search\\
Winston (et al.) & 2009, 2010 & Spitzer/IRAC+Chandra/ACIS & some
potentially confusing source numbers in literature; fixed here\\
Itoh (et al.) & 2010 & optical & spectroscopy of brown dwarf candidates\\
WISE & (2010) & 3.5, 4.6, 12, 22 \mum\ & data obtained in 2010 as part
of All-Sky Survey; AllWISE reduction used, but only in certain very
specific cases (not broadly nor blindly used). \\
Arnold (et al.) & 2012 & MIR & spectroscopy with Spitzer IRS  \\
3XMM-DR4 & 2013 & X-ray (XMM) & all-sky catalog from XMM\\
Foster (et al.) & 2015 & NIR & spectroscopy with APOGEE\\
Sadavoy (et al.) & 2014 & FIR & Class 0 Herschel cross-identifications\\
YSOVAR & 2014 & 3.6, 4.5 \mum & means from YSOVAR monitoring included;
variables not yet recognized as YSOs are included here\\ 
\enddata
\end{deluxetable}

\section{Large Datasets}
\label{sec:largedatasets}

In this section, I discuss the largest area surveys I used; they are
listed, along with the smaller catalogs from the next section, in
Table~\ref{tab:allstudies}.

All of the large-area catalogs described here were merged by position
with a catalog-dependent search radius (usually $\sim$1$\arcsec$).
Many sources, especially those in regions where many sources are close
together on the sky, were individually inspected and matched by hand.
SEDs were constructed as an additional check on the source matching;
objects with particularly strange initial SEDs were also individually
inspected and matched by hand, which often resolved any SED issues.

 \subsection{2MASS and 2MASS 6$\times$}

The Two-Micron All Sky Survey (2MASS; Skrutskie \etal\ 2006) catalog
provided the original nucleus of the catalog, to which all other
catalogs were merged by position, typically with a $\sim$1$\arcsec$
search radius.

NGC 1333 was in the original 2MASS survey, of course, and it was also
included in the long exposure 6$\times$ 2MASS program. The original
2MASS data were obtained in 2000. I included the main 2MASS catalog
and the deeper 6$\times$ catalog near-IR (NIR) $JHK_s$ data into the
database.

\subsection{Cryogenic-era Spitzer Archival Data}

Early in the Spitzer mission, NGC 1333 was observed by both the
guaranteed time observations (GTO) and the original Cores-to-Disks
(c2d) Legacy program (Evans \etal\ 2003, 2009). For both the Infrared
Array Camera (IRAC; Fazio \etal\ 2004) and Multiband Imaging
Photometer for Spitzer (MIPS; Rieke \etal\ 2004) data, the
observations were obtained at multiple epochs separated by at least a
few hours to allow asteroids to move and thus be identified as
asteroids (as opposed to embedded objects in NGC 1333).  The IRAC
observations are at 3.6, 4.5, 5.8, and 8 \mum; the MIPS observations
are at 24, 70, and 160 \mum, but the 160 \mum\ data is very difficult
to interpret in this region and are not included here. The first IRAC
observation was part of GTO program 6, obtained on 2004-02-10; a
second group of IRAC observations was part of c2d, program 178, on
2004-09-08. The first MIPS observations were part of GTO program 58,
obtained on 2004-02-03; an additional 3 epochs were part of c2d
program 178, obtained on 2004-09-19 and 2004-09-20.

As discussed in R14, the cryogenic data were combined and reduced
identically to the YSOVAR monitoring data, except using cryogenic
calibrations and combining the two observations into a single
effective epoch (rather than maintaining separate measurements for
each epoch). The apertures we used were 2.4$\arcsec$. The data were
bandmerged across Spitzer bands by position, and then to the 2MASS
catalog, within a search radius of $1\arcsec$.  

Gutermuth \etal\ (2008a, 2009, 2010) present methodology for
identifying YSOs from the cryogenic catalog. The details of the
selection process appear in those papers, but in summary, multiple
cuts in multiple color-color and color-magnitude diagrams are used to
identify YSO candidates, as distinct from, e.g., extragalactic and
nebular contamination.  This color selection process was run anew on
the re-reduced data.

Spitzer data are also available from the c2d program final data
delivery. The data used for the final delivery are typically the same
BCDs as were used in the cryogenic data that we re-reduced above. As
such, then, they are not independent measurements, and these data were
only used to supplement the cryogenic-era catalog if a band was
missing. There is more information on what I extracted from the c2d
catalogs below.

\subsection{Chandra ACIS}

Chandra X-ray Observatory Advanced CCD Imaging Spectrometer for
wide-field imaging (ACIS-I) observations of NGC 1333 were first
reported in Getman (2002) and then Winston \etal\ (2009, 2010). There
are three pointings in NGC 1333, with obsids 642, 6436, and 6437, with
a total exposure time of 119.3 ks. 

As was described in R14, we re-reduced the Chandra data in a
self-consistent way across most of the YSOVAR clusters. Source
detection was performed using CIAO (Chandra Interactive Analysis of
Observations; Fruscione \etal\ 2006). Sources, even faint ones, were
retained if they had a counterpart in the cryogenic IRAC catalog.
Sources from Chandra were matched to the rest of the catalog with a
position-dependent search radius; see R14. 

Cross-IDs from these X-ray papers are included; see below. However,
the X-ray flux measurement data are included in their entirety in
Rebull \etal\ (2015), the paper on the NGC 1333 YSOVAR data, and do
not play a role in the SEDs, so they are not explicitly included here.
As R14 describes, we identify some candidate cluster members by
looking for objects with star-like SEDs and an X-ray detection. There
are only two new X-ray candidate cluster members that are introduced
as part of this process (J032913.47+312440.7 \& J032837.85+312525.3);
all of the other so-identified members were identified in the
literature as (candidate) members already.

\subsection{WISE}
\label{sec:wise}

The Widefield Infrared Survey Explorer (WISE; Wright \etal\ 2010)
surveyed the whole sky at 3.5, 4.6, 12, and 22 \mum; all of the
available WISE data taken between 2010 Jan and 2011 Feb were
incorporated into the AllWISE catalog. WISE has lower spatial
resolution than Spitzer, and is on average less sensitive. I do not
generally include the AllWISE data, since NGC 1333 is often a
complicated region with high surface brightness, and because we have
extensive higher spatial resolution Spitzer data. However, WISE
provides a band at 12 \mum\ that is not available from Spitzer. I have
incorporated WISE data for certain individual sources, where viable
photometry is available from WISE and the photometry from Spitzer is
incomplete or results in an unusual SED shape. (For example, in some
cases, the [24] point seemed unphysical in the context of the SED, but
the [22] point is well-matched to it.) Comparison of the WISE images
with the Spitzer images was also useful in certain circumstances, such
as for investigating the influence of image artifacts -- artifacts
change between Spitzer and WISE, but sources on the sky should not.

We used the AllWISE data release in nearly all cases, but we now note
an exception. In one very crowded region (sources that are components
of IRAS 7), the WISE All-Sky catalog does a better job of separating
the sources than the AllWISE catalog. AllWISE seems to have inferred
that there was significant proper motion of one of the sources in the
clump, and by inspection of the images, this is not correct. Thus,
WISE flux densities for our catalog sources J032910.70+311820.9 and
J032911.24+311831.8 are taken from the All-Sky catalog, not the
AllWISE catalog.

\section{Literature data and source reconciliation}
\label{sec:litdata}

Many studies have been made specifically of NGC 1333 -- it is one of
the most well-studied star forming regions, with $>$200 refereed
publications in ADS.  It is difficult to include data from every
single paper, especially since so many papers focus on just one or a
few objects in the region, or on just extended objects (e.g., Raga
\etal\ 2012). I endeavored to include in the catalog the most recent
catalogs of point sources and/or those that had the largest
repositories of data, and/or those that were the origin of some source
names still in common use today.  The majority of the information
actually included in the database from the earliest studies is  the
cross-identifications (cross-IDs) with the literature; additional
broadband photometry was included where possible, and not superceded
by subsequent reprocessing of the data.  In the process of assembling
the literature catalog of sources, I reconciled many ambiguities and
inconsistencies in the literature. I provide below descriptions of the
more complicated issues. All of the reconciled cross-identifications
in NGC 1333 are included in Table~\ref{tab:crossids}, including an
indication of whether or not the corresponding survey identified the
object as a YSO.  

All of the aggregate $J$-[24] single-epoch photometry appears in
Table~\ref{tab:bigdata}. There are nearly 7000 objects in the catalog,
about 300 of which are identified in the literature as YSO candidates.

IAU standards recommend not renaming previously-identified sources,
but as one of the purposes of the present catalog is to sort out
inconsistencies and inacuracies in existing catalogs, assigning a new,
coordinate-based name seems appropriate. The coordinate based names
presented here (in Tables~\ref{tab:crossids} and \ref{tab:bigdata})
should be preceeded by `R15-NGC1333.' 

\begin{figure}[ht]
\epsscale{1.0}
\plotone{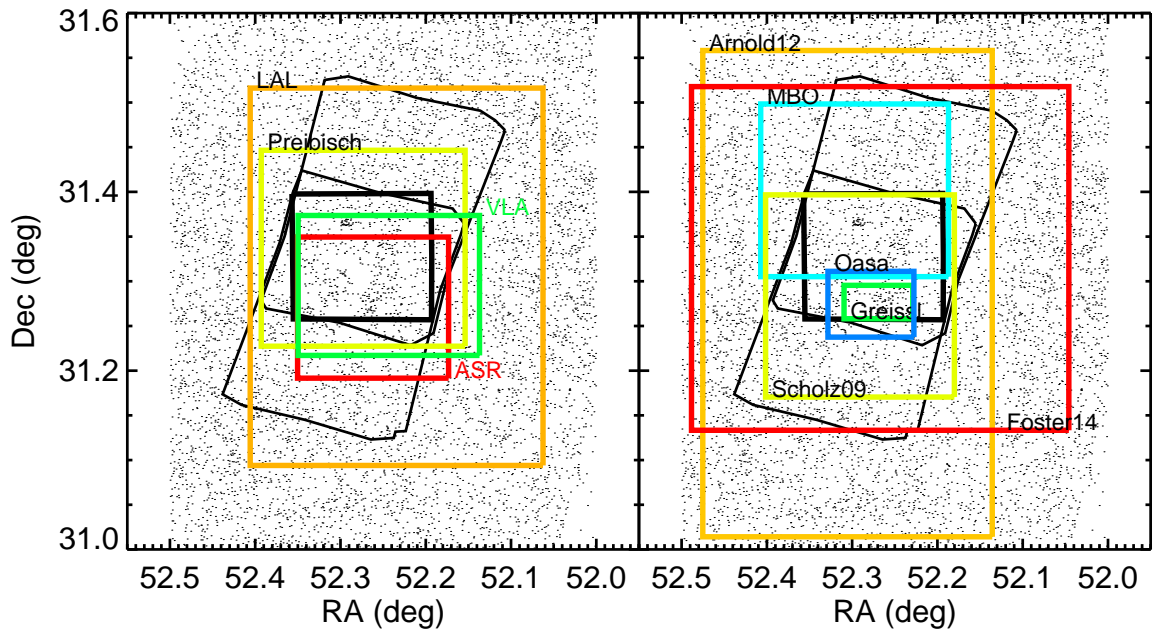}
\caption{Approximate spatial coverage of the various catalogs included
in this survey, distributed over two plots simply for clarity. Each
small  black dot corresponds to an object in the final catalog
assembled here. The black polygons in both plots are the coverage of
the 3.6 and 4.5 \mum\ YSOVAR monitoring regions, and the thick black
square in both plots is the Chandra coverage. Colored squares
correspond roughly to most of the other surveys included here, as
indicated in the corners of the squares. For these colored squares,
note that what is shown is the min/max of the RA/Dec; no attempt is
made at polygon representation of the other survey regions, because
this figure is simply meant to give a rough indication of the regions
of the cluster observed in these respective surveys, e.g., a
relatively small region has been probed with NICMOS (the green Greissl
\etal\ square).  }
\label{fig:wherecat}
\end{figure}

Figure~\ref{fig:wherecat} gives a rough indication of the various
larger surveys included here. The footprints from Chandra and from the
YSOVAR monitoring are shown close to their actual coverage. No attempt
is made to capture complex polygons of coverage for the other surveys,
but just squares encompassing all sources are shown; in other words,
there are no data from a given survey outside of its square, but there
may be incomplete spatial coverage inside of it.  

\subsection{On Coordinate Accuracies}
\label{sec:coord}

Integral to the process of source matching across catalogs is an
understanding of the systematic and statistical errors present in the
positions of the objects in the catalogs. 2MASS provides a very high
quality coordinate system, $<$0.1$\arcsec$ with respect to the 
International Celestial Reference System (ICRS) reference frame for
bright sources, over the whole sky. WISE and Spitzer coordinates are
fundamentally tied to this 2MASS coordinate system. WISE positional
uncertainties are typically $<$0.2$\arcsec$, often much less.
Spitzer/IRAC positional uncertainties are comparable at
$<$0.2$\arcsec$.

I present the prior studies I integrated in this section in roughly
chronological order. However, in practice, I iteratively (and very
often manually) compared each of the sources in the older catalogs to
the 2MASS catalog and images, adjusting or correcting coordinates as
needed, before merging all the catalogs together. In many cases, it
was a simple shift within $\lesssim$5$\arcsec$ from a place on the sky
without a 2MASS source to the location of a relatively bright 2MASS
source.  I note that the Infrared Science Archive (IRSA) tool
FinderChart\footnote{http://irsa.ipac.caltech.edu/applications/finderchart/}
was extremely helpful for this process. FinderChart uses data from
WISE, 2MASS, and the Digitized Sky Survey (DSS), which is a
digitization of the photographic sky survey plates from the Palomar
(Palomar Observatory Sky Survey) and UK Schmidt telescopes.
FinderChart provides a thumbnail image of the sky at multiple
wavelengths, making comparisons relatively straightforward.

For any catalog, the coordinate accuracy depends on there being
sufficient numbers of point sources in the field of view (or mosaic)
in order to anchor the coordinate system. This has two ramifications
for the catalogs considered in this section. First, for some early NIR
observations, arrays were very small (for example, Aspin \etal\ 1994
had a single pointing field of view of just under an arcminute on a
side), resulting in relatively few point sources per pointing, making
astrometry very hard to do accurately. Moreover, then as now,
astrometry was bootstrapped to prior observations, but at that time, this
bootstrapping had to occur without the reliable all-sky anchor
provided by 2MASS. Secondly, for long-wavelength observations such as
Spitzer/MIPS and Herschel maps, there are many fewer point sources
that can be linked to the 2MASS coordinate system, so the positional
uncertainty can be worse, and sometimes astrometry relative to one or
a few point sources is the best available.  It can be difficult to
make a clear correspondence between the shorter wavelength and longer
wavelength sources, not just because of the coordinates, but also
because the emission may not be coming from the same location in/near
the object. I have attempted to make these matches here, being mindful
of the fact that these long wavelength sources are often in regions
that are extremely complex, with high surface brightness (complicating
both photometry and astrometry). Moreover, the source of the long
wavelength emission may just be by chance aligned with an emitter of
short wavelength emission. I believe what I have done is correct, but
I have provided descriptions below of what I have done in the event
that subsequent investigators disagree.

Another source of positional uncertainty is the space motions of these
obejcts.  Over the time baselines considered here, could the sources
in NGC 1333 be moving significantly? Karchenko \etal\ (2013) report
that for NGC 1333 the average $\mu_{\rm RA}, \mu_{\rm
Dec}$=5.51,$-$10.28 mas yr$^{-1}$, so over 50 years (longer than
considered here), a typical object could move $\sim$0.5$\arcsec$.
However, the spatial resolution for both 2MASS and IRAC are both
$\sim$1$\arcsec$, and these catalogs are the input catalogs that
establish the bulk of the master NGC 1333 catalog. I search for
counterparts between catalogs with a typical tolerance of
$\sim$1$\arcsec$. Therefore, most sources will not have enough proper
motion over the baselines considered here to affect our catalog
merging. Assessing the motion of these objects based on these data is
theoretically possible, but beyond the scope of this work.

\begin{deluxetable}{ccccp{7cm}}
\tabletypesize{\scriptsize}
\tablecaption{Contents of Cross-ID Catalog\tablenotemark{a}\label{tab:crossids}}
\tablewidth{0pt}
\tablehead{\colhead{Bytes} & \colhead{Format)}
&\colhead{Units} & \colhead{Label} & \colhead{Explanations}}
\startdata
   1- 18& A18 &  --- &    cat       & Catalog name (HHMMSS.ss+DDMMSS.s; J2000); should be preceeded by `R15-NGC1333 J'                                   \\
  19- 23& I5  &  --- &    ASR       & ASR number                                                                \\
  24- 27& I4  &  --- &    ASR YSO?  & Did ASR tag it as a YSO? 1=yes, 0=no, $-$9 no information                    \\
  28- 32& I5  &  --- &    LAL       & LAL number                                                                \\
  33- 36& I4  &  --- &    LAL YSO?  & Did LAL tag it as a YSO? 1=yes, 0=no, $-$9 no information                    \\
  37- 41& I5  &  --- &    VLA       & VLA number                                                                \\
  42- 45& I4  &  --- &    VLA YSO?  & Did Rodriguez et al. tag it as a YSO? 1=yes, 0=no, $-$9 no information       \\
  46- 50& I5  &  --- &    Preibisch & Preibish number                                                           \\
  51- 70& A20 &  --- &    2MASS num & 2MASS name                                                                \\
  71- 75& I5  &  --- &    Getman num& Getman number                                                             \\
  76- 79& I4  &  --- &    Getman var& Did Getman tag it as {\em variable}? 1=yes, 0=no, $-$9 no information              \\
  80- 84& I5  &  --- &    MBO num   & MBO number                                                                \\
  85-103& A19 &  --- &    c2d num   & c2d name                                                                  \\
 104-107& I4  &  --- &    c2d YSO?  & Did c2d tag it as a YSO? 1=yes, 0=no, $-$9 no information                    \\
 108-117& A10 &  --- &    Greissl num& Greissl \etal\ (2007) number                                             \\
 118-123& A6  &  --- &    Hatchell num&  Number from Hatchell \etal\  (2007)                                    \\
 124-128& I5  &  --- &    Oasa num  & Number from Oasa \etal\ (2008) number                                     \\
 129-132& I4  &  --- &    Oasa YSO? & Did Oasa \etal\ tag it as a member? 1=yes, 0=no, $-$9 no information         \\
 133-137& I5  &  --- &    G08 num   & Number from Gutermuth \etal\ (2008)                                       \\
 138-141& I4  &  --- &    G08 YSO?  & Did Gutermuth \etal\ tag it as a member? 1=yes, 0=no, $-$9 no information   \\
 142-146& I5  &  --- &    S09 num   & Number from Scholz \etal\ (2009)                                          \\
 147-151& I5  &  --- &    Winston nun&  Number from Winson \etal\ (2010)                                         \\
 152-161& A10 &  --- &    S12 num   & Number from Scholz \etal\ (2012)                                          \\
 162-165& I4  &  --- &    S12 YSO?  & Did Scholz \etal\ tag it as a member? 1=yes, 0=no, $-$9 no information       \\
 166-170& I5  &  --- &    A12 num   & number from Arnold \etal\ (2012)                                          \\
 171-174& I4  &  --- &    A12 YSO?  & Did Arnold \etal\  tag it as a YSO? 1=yes, 0=no, $-$9 no information         \\
 175-179& I5  &  --- &    Foster num& number from Foster \etal\ (2015)                                          \\
 180-207& A19 &  --- &    YSOVAR    & YSOVAR name from Rebull \etal\ (2015)                                     \\
 208-211& I4  &  --- &    YSOVAR YSO?& Did Rebull \etal\  tag it as a YSO? 1=yes, 0=no, $-$9 no information        \\
 212-317& A105 &  --- &    other     & Any other names or cautions for this object                                           \\
\enddata
\tablenotetext{a}{Entire data table available online at the journal.}
\end{deluxetable}

\begin{deluxetable}{ccccp{6cm}}
\tabletypesize{\scriptsize}
\tablecaption{Contents of Single-Epoch 1-70\mum\ Catalog\tablenotemark{a}\label{tab:bigdata}}
\tablewidth{0pt}
\tablehead{\colhead{Bytes} & \colhead{Format)}
&\colhead{Units} & \colhead{Label} & \colhead{Explanations}}
\startdata
   1- 18 &A18   &--- &    cat       &Catalog name (HHMMSS.ss+DDMMSS.s; J2000)           \\ 
  19- 28 &F10.6 &deg &    RA        &RA, J2000, decimal degrees                         \\ 
  29- 38 &F10.6 &deg &    Dec       &Dec, J2000, decimal degrees                        \\ 
  39- 42 &A4    &--- &    l\_J       &Limit flag on J                                    \\ 
  43- 49 &F6.2  &mag &    J         &J band magnitude                                   \\ 
  50- 55 &F6.2  &mag &    e\_J       &Uncertainty in J                                   \\ 
  56- 59 &A4    &--- &    l\_H       &Limit flag on H                                    \\ 
  60- 66 &F6.2  &mag &    H         &H band magnitude                                   \\ 
  67- 72 &F6.2  &mag &    e\_H       &Uncertainty in H                                   \\ 
  73- 76 &A4    &--- &    l\_Ks      &Limit flag on Ks                                   \\ 
  77- 83 &F6.2  &mag &    Ks        &Ks band magnitude                                  \\ 
  84- 89 &F6.2  &mag &    e\_Ks      &Uncertainty in Ks                                  \\ 
  90- 93 &A4    &--- &    l\_[3.6]   &Limit flag on [3.6]                                \\ 
  94-100 &F6.2  &mag &    [3.6]     &Spitzer/IRAC 3.6um band magnitude               \\ 
 101-106 &F6.2  &mag &    e\_[3.6]   &Uncertainty in [3.6]                               \\ 
 107-110 &A4    &--- &    l\_[4.5]   &Limit flag on [4.5]                                \\ 
 111-117 &F6.2  &mag &    [4.5]     &Spitzer/IRAC 4.5um band magnitude               \\ 
 118-123 &F6.2  &mag &    e\_[4.5]   &Uncertainty in [4.5]                               \\ 
 124-127 &A4    &--- &    l\_[5.8]   &Limit flag on [5.8]                                \\ 
 128-134 &F6.2  &mag &    [5.8]     & Spitzer/IRAC 5.8um band magnitude              \\ 
 135-140 &F6.2  &mag &    e\_[5.8]   &Uncertainty in [5.8]                               \\ 
 141-144 &A4    &--- &    l\_[8.0]   &Limit flag on [8.0]                                \\ 
 145-151 &F6.2  &mag &    [8.0]     &Spitzer/IRAC 8.0um band magnitude               \\ 
 152-157 &F6.2  &mag &    e\_[8.0]   &Uncertainty in [8]                                 \\ 
 158-161 &A4    &--- &    l\_[24]    &Limit flag on [24]                                 \\ 
 162-168 &F6.2  &mag &    [24]      &Spitzer/MIPS 24um band magnitude               \\ 
 169-174 &F6.2  &mag &    e\_[24]    &Uncertainty in [24]                                \\ 
 175-178 &A4    &--- &    l\_[70]    &Limit flag on [70]                                 \\ 
 179-185 &F6.2  &mag &    [70]      & Spitzer/MIPS 70um band magnitude              \\ 
 186-191 &F6.2  &mag &    e\_[70]    &Uncertainty in [70]                                \\ 
 192-204 &A12   &--- &    SpTy      & Spectral type                                     \\ 
 205-215 &A10   &--- &    SpTySrc   & Origin of spectral type (literature)              \\ 
 216-221 &I4    &K   &    Teff      & \teff\ from Foster \etal\ (2015)                    \\ 
 222-227 &F6.2  &--- &    oursedslope24&  our SED slope from 2 to 24
 \mum\ if YSO candidate\\ 
 228-233 &A6    &--- &    oursedclass24&  our SED class using the
 slope from 2 to 24 \mum\ if YSO candidate, else 'notY' \\ 
 234-239 &F6.2  &--- &    oursedslope8 & our SED slope from 2 to 8 
 \mum\ if YSO candidate \\
 240-245 &A6    &--- &    oursedclass8 & our SED class using the slope
 from 2 to 8 \mum\ if YSO candidate, else 'notY'   \\ 
 246-252 &A7    &--- &    G08class  & SED class from Gutermuth \etal\ (2008)            \\ 
 253-259 &A7    &--- &    A12class  & SED class from Arnold \etal\ (2012)               \\ 
 260-266 &A7    &--- &    H07class  & SED class from Hatchell \etal\ (2007)             \\ 
 267-274 &A7    &--- &    S14class  & SED class from Sadavoy \etal\ (2014)              \\ 
\enddata
\tablenotetext{a}{Entire data table available online at the journal.}
\end{deluxetable}

\clearpage

\subsection{ASR Catalog}
\label{sec:appasr}

\begin{figure}[ht]
\epsscale{1.0}
\plotone{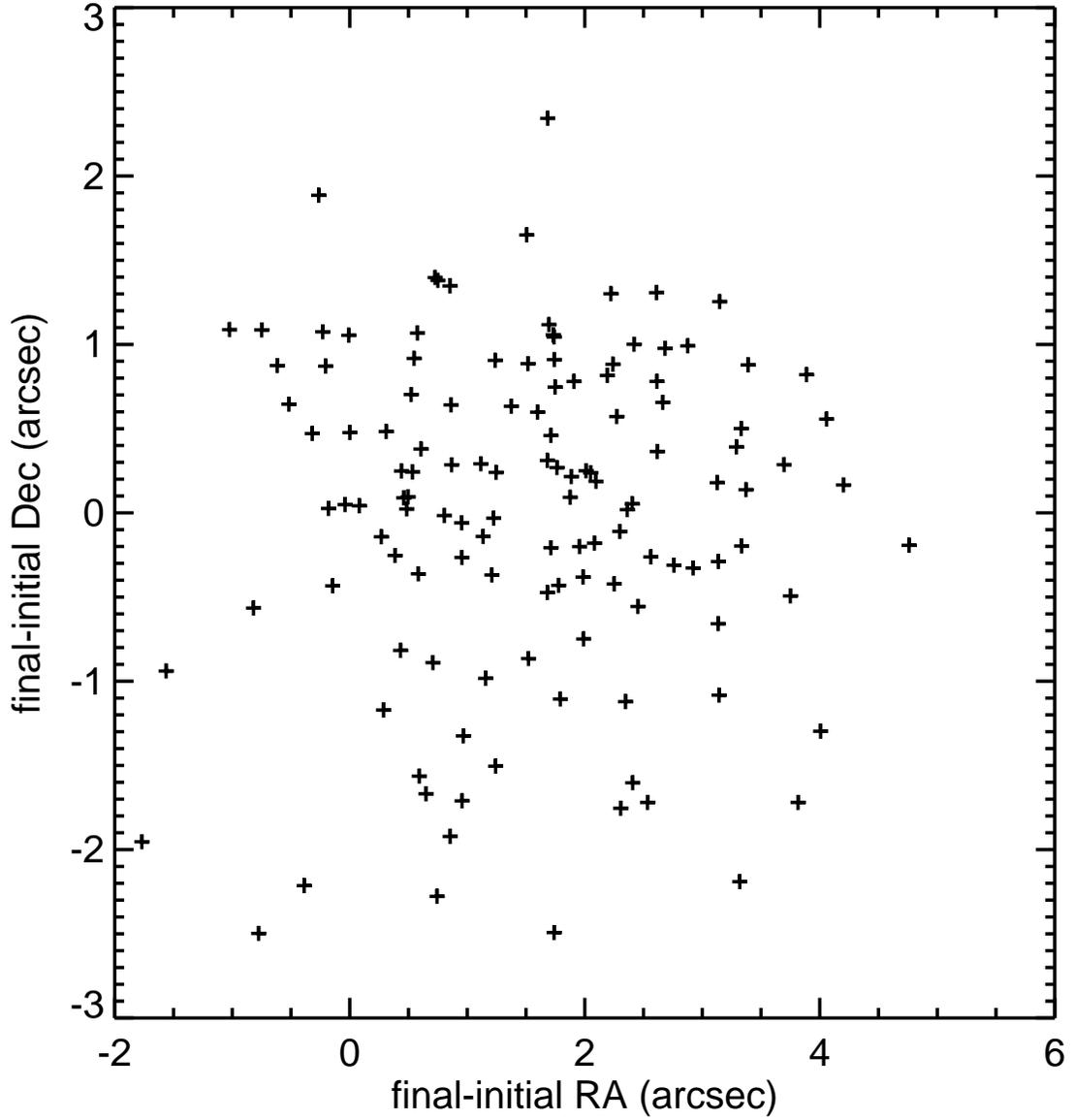}
\caption{Comparison of offsets in RA and Dec (both in arcseconds)
between the final, adopted position (matching 2MASS) and the intitial
position precessed from the B1950 coordinates provided in ASR. The
offests are symmetric in declination, and reveal a systematic offset
in RA. The median net offset is $\sim$1.7$\arcsec$. }
\label{fig:asrposuncert}
\end{figure}

\begin{figure}[ht]
\epsscale{1.0}
\plotone{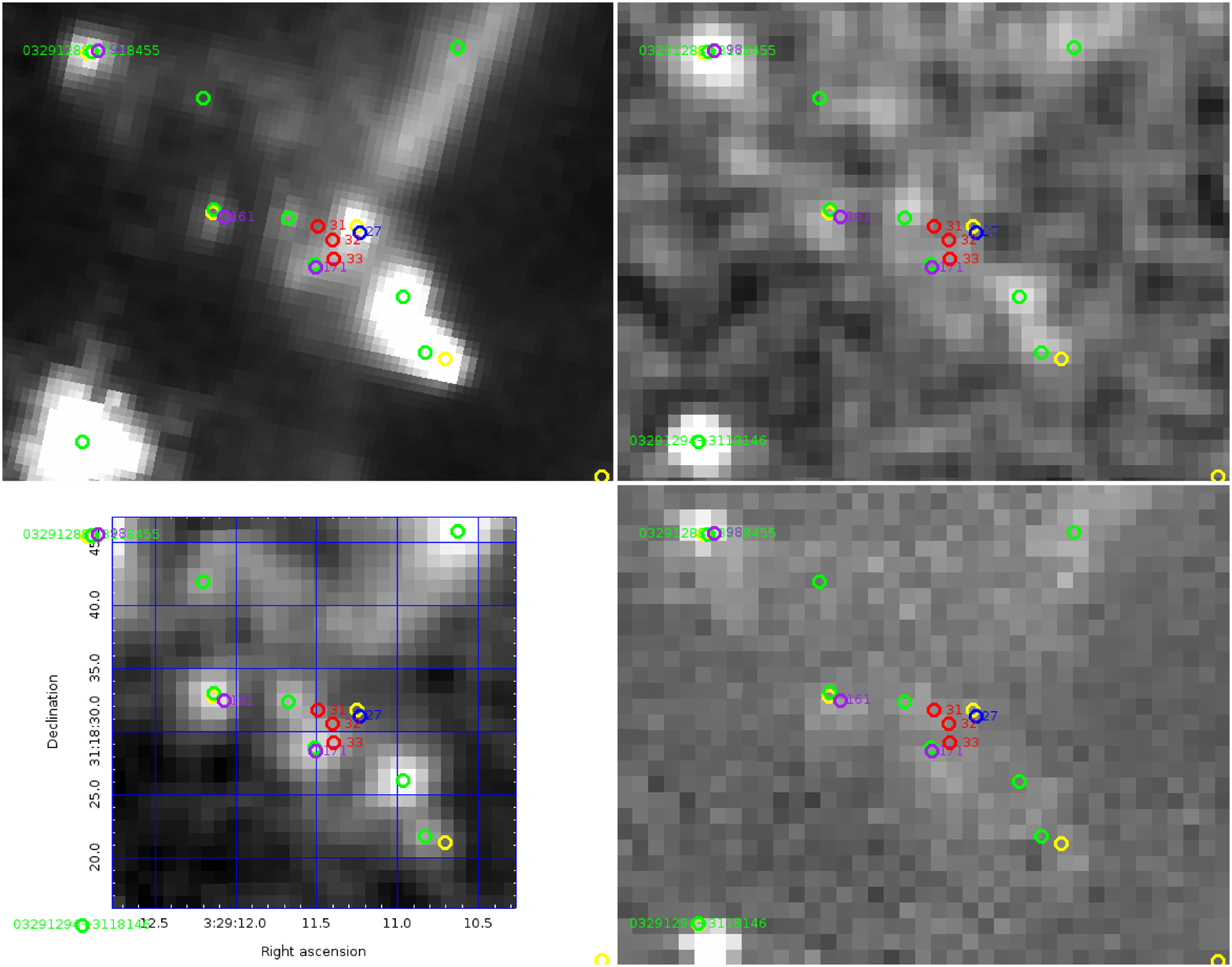}
\caption{Resolving the ASR 31-32-33 confusion. The images shown are
(left to right, top to bottom): IRAC-1 (3.6 \mum) from the cryogenic
mission, 2MASS $K_s$ from the classic survey, 2MASS $K_s$ from the
6$\times$ deeper survey, and then the original ASR image from 1994,
unfortunately with subsequently updated astrometry. The  sky is shown
on the same scale  in each image; RA and Dec coordinates are indicated
in the lower left for reference. Sources are overlaid as follows: red
circles are ASR 31/32/33 (ordered top to bottom), with the original
coordinates as reported and simply precessed to J2000 from 1950; green
circles: 2MASS sources, where ones that have the long names indicated
are from 2MASS classic and ones without numbers are from 2MASS
6$\times$; yellow circles: full cryogenic catalog from Gutermuth
\etal\ (2008); purple circles: Wilking \etal\ (2004) MBO catalog; 
blue circle: Rodriguez \etal\ (1999) catalog (VLA 27).  See text for
detailed discussion of which sources are identified with each other in
the ASR 31-32-33 region.}
\label{fig:ASR313233}
\end{figure}

Aspin \etal\ (1994; ASR) was one of the first wide-field NIR surveys
of NGC 1333 reported in the literature. They published a full $JHK$
catalog (i.e., not just the objects they thought were cluster members)
for 134 objects to $K=16.2$. I incorporated their full catalog into
our database, keeping track of the objects tagged by ASR as likely
YSOs. Numbers from this catalog are preceeded by ``ASR.'' $JHK$
measurements from this catalog were not retained because there were
2MASS measurements available. Moreover, the photometry on average is
well-matched, typically within $\sim$0.2 mag (often less); while most
of the stars in the catalog are unlikely to be members, young stars
are expected to be intrinsically variable at levels greater than a few
tenths of a magnitude.

Since this is the first (and oldest) literature catalog I consider
here, I report some additional information on it as representative of
the difficulties inherent in this process, and my approach to it.
While an individual field of view was just under an arcminute on a
side, ASR mapped $\sim$10$\arcmin \times$10$\arcmin$. They established
astrometry by comparison to Herbig \& Jones (1983) for 30 objects,
estimating uncertainties of $\sim$1$\arcsec$ as a result.  The
original ASR catalog was reported in B1950 coordinates, which I
precessed and overlaid on the images and compared to the 2MASS and
IRAC catalogs. ASR was a relatively shallow $K$-band observation, 
comparable to 2MASS classic (as opposed to 2MASS 6$\times$). These
sources (particularly the NGC 1333 members) are intrinsically
variable. However, while I expected that variability at $K$ could make
the object appear or disappear out of a relatively shallow survey, I 
still expected it to show up at the 2MASS 6$\times$ depth or at least
the IRAC maps, not necessarily gone entirely.  The IRAC observations
reach at the very least the same sources as 2MASS, for most SEDs; even
relatively shallow IRAC observations commonly reach sources fainter
than are detected by 2MASS. For nearly 90\% of the 134 ASR sources, I
could find counterparts within 5$\arcsec$ in the 2MASS catalog, with a
median positional offset of $\sim$1.7$\arcsec$, and larger offsets in
RA than in Dec; see Figure~\ref{fig:asrposuncert}. For most of the
sources, particularly the larger positional offsets, images from 2MASS
were compared to the positions provided in the original ASR to ensure
that the match was appropriate. In most cases, the match was readily
apparent.  

Unlike most of the rest of the cluster, the region containing the
trio of ASR 31/32/33 is quite confusing, as can be seen in
Figure~\ref{fig:ASR313233}. I now discuss this region in detail.

ASR 33 is the easiest of this triplet, and is probably the same as MBO
171 (which is the same cross-identification provided by Wilking \etal\
2004). It is catalog source R15-NGC1333 J032911.51+311828.6.

VLA 27 has been tied (including by myself in Rebull \etal\ 2003) to
one of the ASR 31/32/33 sources. But note that there is an IRAC source
coincident with the position reported for VLA 27, offset (by
$\sim$4.5$\arcsec$, a significant amount by IRAC standards) from ASR
31/32/33. And, note that there is no 2MASS source (in the shallower or
deeper 2MASS data).  The original ASR survey could not have been
enough deeper than 2MASS 6$\times$ to detect a source not recovered by
either of the 2MASS surveys; intrinsic variations of this level in
this source are possible but unlikely. I have tied VLA 27 to the IRAC
source (R15-NGC1333 J032700.47+313725.9), but not the ASR sources. The
VLA positions should be very good, so leaving this source affiliated
with the IRAC source with which it is coincident seems appropriate.

There is a rough arc of 2MASS sources with the same rough concavity as
the arc of ASR 31/32/33 sources. But there are 5 total 2MASS sources,
and only three ASR sources. Could the northern three 2MASS sources be
the three ASR sources?  Wilking \etal\ (2004), a.k.a. the MBO source
numbers below, linked ASR 31 to MBO 161. Given the catalog sources
overlaid in Fig.~\ref{fig:ASR313233}, MBO 161 is {\em not} the same as
ASR 31.

The original ASR astrometry has some uncertainties, but not as bad as
would be required to enforce the match of the two arcs of sources,
especially over this small region. I obtained the original data from
ASR (C. Aspin, priv.\ comm.), which had been updated
with more recent astrometry; see Figure~\ref{fig:ASR313233}. There are
two sources in the left corner of the ASR data bright enough to be in
2MASS classic. Note that the originally reported MBO, IRAC, and 2MASS
source positions are a little to the south of the northern source, and
a little to the north of the southern source, which should be
representative of the net positional errors in this immediate region.
The original ASR coordinates in this region also reflect this tiny
distortion in the North-South direction, less so in the East-West
direction. The local astrometry is probably correct (unless there is a
tile boundary here, which does not seem to be the case in the ASR
image; see also LAL coordinate matching below). So the differential
astrometry should be high-quality among the local ASR sources. There
is nothing approaching the magnitude of the distortions that would be
required to map ASR 31 onto MBO 161, and ASR 32 onto the 2MASS source
that would be in between the two other ostensible ASR matches.

I conclude that:
\begin{itemize}
\item ASR 31 should be matched to catalog source J032911.67+311832.3
(and c2d source J032911.6+311832).
\item ASR 32 is likely spurious and cannot be tied to any source
detected in these other surveys to date, so I have removed it from the
catalog. It may be that it should be matched to VLA 27 (others have
made this match in the literature), but I have not enforced this match
here.
\item ASR 33 should be matched to MBO 171 and catalog source
J032911.51+311828.6 (and c2d source J032911.5+311828).
\item VLA 27 should match to catalog source J032911.24+311831.8 (and
c2d source J032911.3+311831).
\item ASR 31 should {\em not} be matched to MBO 161 (which is
different than what is found in the literature). MBO 161 is matched to
the catalog source J032912.13+311832.1 (and c2d source
J032912.1+311832).
\end{itemize}

Another more minor ASR issue pertains to ASR 50, which is clearly in
the original ASR image, and can be seen in 2MASS 6$\times$, but is on
a diffraction spike and resembles an artifact enough that it was
probably dropped by the 2MASS pipeline as a result. There is an IRAC
source within 2.5$\arcsec$ of the Aspin source (J032855.08+311416.4)
that is most likely the match.

There are several remaining ASR sources for which 2MASS-classic
counterparts cannot easily be found. ASR 5=J032904.53+311554.6, oddly,
can be seen clearly in the DSS images, but by 2MASS $JHK_s$ is
extremely faint, and is lost in the glare of a nearby very bright
source by Spitzer and WISE bands. I have retained it with the
original ASR coordinates, since it can be clearly identified in the
DSS and is well-matched to that position. ASR 20, 21, and 22 are all
in a region that becomes bright with extended emission at Spitzer
bands. ASR 20 does not have a clear match in 2MASS, but ASR 21 is
faintly seen in the 2MASS $K_s$ images, and may be extended. There are
two sources from the 6$\times$ catalog (both of which may be part of
the extended emission); ASR 21 is matched to the closer one. At 3.6
\mum, ASR 22 is close to a blob of extended emission which may have
concentrated knots within it; I have tied it to 2MASS
J03290842+3115284 as the brightest source in the blob. ASR 71 is
entirely within another region that, at 3.6 \mum, is a blob of
extended emission that may have concentrated knots within it. The
brightest portion of the blob is already identified with ASR 49. It is
unclear to what object at 2MASS bands ASR 71 should be matched, so I
have retained it with the original ASR coordinates.  ASR 119 is almost
exactly in between two 2MASS sources that are very far away, one
$\sim$15$\arcsec$ north, and one $\sim$18$\arcsec$ south. Lacking a
compelling reason to look this far away for a counterpart and make the
association with one or the other, I have left ASR 119 alone at its
original location despite having no counterpart. ASR 133 is similiarly
more or less in between two sources $\sim$15$\arcsec$ north and south
of this position; similarly, I have left it alone.  ASR 134 does not
have a 2MASS-classic counterpart, but can be seen faintly on the $K_s$
images, and it is bright by IRAC bands.  ASR 50, 75, 94, and 98 all
have IRAC matches within 5$\arcsec$, even if counterparts are not
apparent in the 2MASS-classic images.

\subsection{LAL Catalog}
\label{sec:applal}

\begin{figure}[ht]
\epsscale{1.0}
\plotone{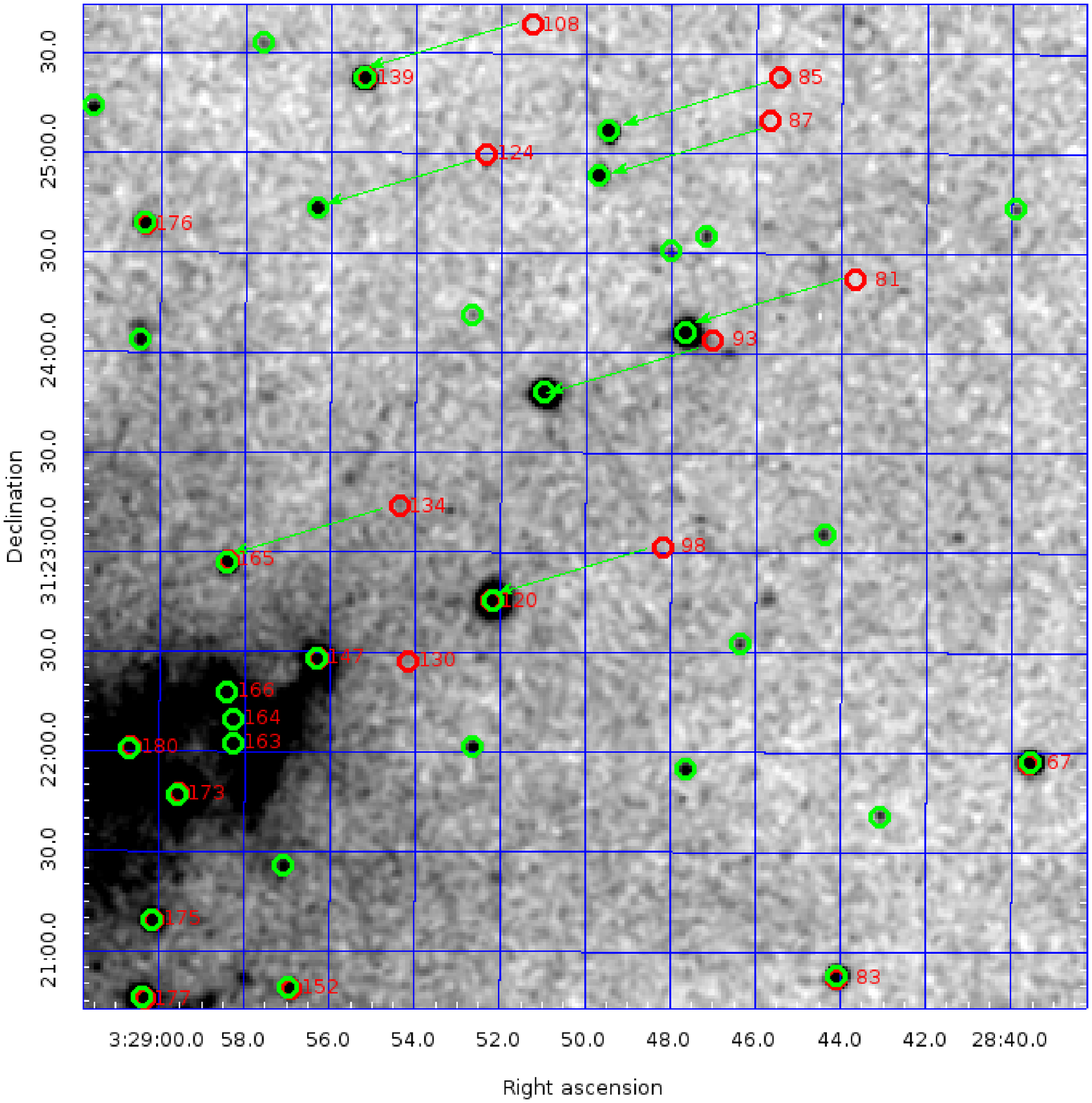}
\caption{Resolving some of the LAL coordinate issues. The background image
is an inverse scale 2MASS $K_s$-band image. In the image, the red
circles with numbers are the original LAL coords as reported; the
green boxes are sources in 2MASS (classic and 6$\times$).  Essentially all of
the clear point sources seen in this 2MASS image are recovered by the
2MASS catalog, as expected. The pattern suggested by the set of LAL
sources (81, 85, 87, 93, 98, 108, 124, 134) suggest that the
astrometry in this specific region of the LAL map is offset by
54$\arcsec$; see text for more details. I cannot recover LAL 130, as
an offset by the same direction and magnitude does not land on a 2MASS
source. }
\label{fig:Lal81etc}
\end{figure}

\begin{deluxetable}{cp{15cm}}
\tabletypesize{\scriptsize}
\tablecaption{Summary of LAL Modifications\label{tab:lalmods}}
\tablehead{\colhead{LAL number} & \colhead{notes}}
\startdata
27 & nothing at this location and nearby sources already have matches; removed\\
35 & nothing at this location and nearby sources already have matches; removed\\
39 & nothing at this location and nearby sources already have matches; nearby bright source 2M J03283695+3123121 does not have match, but that is nearly an arcminute away and thus is an unlikely match; removed\\
44 & nothing at this location and nearby source already has match; removed\\
47 & nothing at this location, no nearby sources; removed\\
50 & nothing at this location, and nearby sources already have matches; removed\\
52 & this one, plus 59 and 64, make an arc of 3 sources; this shape is not matched to anything in the 2MASS or IRAC images, and nearby sources have matches; removed\\
59 & see 52; removed\\
61 & near a faint (at 3.6 \mum) extended source in [3.4] (WISE) and [3.6], but not in 2MASS; morphology suggests jet; removed\\
64 & see 52; removed\\
71 & nothing at this location, no nearby sources; removed\\
81 & one of the ``54$\arcsec$ to the southeast'', now 2MASS J03284764+3124061\\
84 & nothing at this location, no nearby sources; removed \\
85 & one of the ``54$\arcsec$ to the southeast'', now 2MASS J03284947+3125066\\
87 & one of the ``54$\arcsec$ to the southeast'', now 2MASS J03284971+3124534\\
93 & 81 moved close to original position of this one as part of the ``54$\arcsec$ to the southeast''; this one moves from 2MASS J03284764+3124061 (where it is offset from a bright star, and to which it has sometimes been tied in the literature) to 2MASS J03285097+3123479\\
98 & one of the ``54$\arcsec$ to the southeast'', now identical to LAL 120, matched to 2MASS J03285216+3122453\\
108 & one of the ``54$\arcsec$ to the southeast'', now identical to LAL 139, matched to 2MASS J03285521+3125223 \\
116 & nothing at this location, and nearby sources already have matches; removed\\
118 & nothing at this location, and several nearby sources all already have matches; removed\\
120 & could justifiably move to match LAL 147, but left it matched to 2MASS J03285216+3122453 (see ``54$\arcsec$ to the southeast'' discussion); no net changes\\
122 & nothing at this location, and nearby sources already have matches; removed \\
124 & one of the ``54$\arcsec$ to the southeast'', now 2MASS J03285630+3124432 \\
130 & nothing at this location, nor at a place offset in the same direction and size as the ``54$\arcsec$ to the southeast''; removed\\
134 & one of the ``54$\arcsec$ to the southeast'', now identical to LAL 165, matched to 2MASS J03285842+3122567\\
135 & nothing at this location, and nearby sources already have matches; removed\\
149 & in original LAL catalog, within 0.6$\arcsec$ of 148, though photometry is different; retained 148 (and removed this one)\\
174 & nothing at this location, and nearby sources already have matches; removed\\
242 & in original LAL catalog, within 0.13$\arcsec$ of 241, and only band available ($K$) matches to 0.02 mag; retained 241 (and removed this one)\\
244 & nothing at this specific location, but on edge of complex extended region with many sources; unclear what should match it; removed \\
259 & in original LAL catalog, 258 and 259 have identical positions and matching photometry; retained 258 (and removed this one)\\
271 & 270-271 are 1.03$\arcsec$ apart, and are likely two components of an object that is multiple in the 2MASS catalog, but both of the LAL sources are much closer to 2M J03291433+3114441 than 2M J03291409+3114423; combined and 270 retained\\
281 & nothing at this location and nearby source already has match; removed\\ 
284 & nothing at this location and nearby sources already have matches; removed\\
289 & in original LAL catalog, 289 and 290 have identical positions though different photometry (one is missing $J$,$H$; $K$ matches to 0.02 mag) identical to 290; retained 290 (and removed this one)\\
291 & nothing at this location and nearby sources already have matches; removed\\
338 & nothing at this location and nearby source already has match; removed\\ 
352 & nothing at this location and nearby sources already have matches; removed\\
\enddata
\end{deluxetable}

Lada \etal\ (1996; LAL) also imaged NGC 1333 in $JHK$, over a region 4
times larger in area than ASR. Their entire catalog was not included
in the original article, but was obtained via private communication
from C.\ Lada. Objects LAL thought were young were also indicated in
this catalog.  Numbers from this catalog are preceeded by ``LAL.''
$JHK$ measurements from this catalog were not retained because 2MASS
measurements were available for all of the retained LAL sources. As
for ASR above, the ensemble of $JHK$ photometry is reasonably
well-matched to 2MASS (better than $\sim$0.2 mag), though there can be
excursions for individual sources. 

The observations presented in LAL were assembled from many smaller
pointings in the NIR, though unlike ASR, a single pointing field of
view for LAL was relatively large, and at $\sim$5$\arcmin$, comparable
to a single IRAC field of view. LAL report that their astrometry is
based on 5 stars from the Hubble Space Telescope Guide Stars Catalog,
and that their estimated positional uncertainty is
0.5$\pm$0.2$\arcsec$.  However, based on my comparison to 2MASS, I
strongly suspect that there are several astrometric problems. In most
cases, the sources are recoverable; for 86\% of the sources, I can
find a counterpart in 2MASS within 3$\arcsec$. However, in some cases,
I could not find a counterpart. In one region, I noticed a pattern
offset which I now describe. 

As seen in Figure~\ref{fig:Lal81etc}, I strongly suspect, based on
relative positions, that several sources in the LAL catalog should all
move 54$\arcsec$ to the southeast.  These sources all appear, perhaps,
to be in the corner of a component tile of the LAL final mosaic, and
evidently the astrometry in this region (tile corner?) was not
well-constrained.  The consistency in the pattern of sources on the
sky is compelling evidence that the sources should be shifted; a
direct comparison of the $JHK_s$ values on a source-by-source basis
often supports this assertion. 

The repercussions are that:
\begin{itemize}
\item LAL 81 moves from having no match to 2MASS J03284764+3124061,
which is nominally close to the original LAL 93 position. There is a
faint source that appears only at 3.6 \mum\ under this
originally-reported location of LAL 81, but the SED for that object is
different from the LAL-reported $JHK$ by many orders of magnitude,
which was what originally led me to suspect that this was not the
correct match. The LAL-reported $JHK$ matches the  2MASS photometry
reasonably well, being offset by 0.16 mag, 0.12, and 0.02 mag in $J$,
$H$, and $K$, respectively.
\item LAL 85 moves from having no match to 2MASS
J03284947+3125066. The photometry matches between these sources to
0.14 mag in $J$, 0.08 mag in $H$, and 0.01 mag in $K$.
\item LAL 87 moves from having no match to 2MASS
J03284971+3124534. The magnitudes match to 0.24, 0.10, and 0.06 mag in
$J$, $H$, and $K$.
\item LAL 93 moves from 2MASS J03284764+3124061 (where it is offset
from a bright star, and to which it has sometimes been tied in the
literature) to 2MASS J03285097+3123479. The photometry matches this
latter source, 2MASS J03285097+3123479, better (0.16, 0.04, 0.01 in
$JHK$) than the former source, 2MASS J03284764+3124061 (0.17, 0.22,
0.41 in $JHK$).
\item There is nothing at the nominal location of LAL 98. If I apply
the  offset in the same direction as the other sources above, then LAL
98 is identical to LAL 120, which is already matched to 2MASS
J03285216+3122453. However, the LAL reported magnitudes for 98 and 120
differ by $\sim$1 mag at each band. On the other hand, 2MASS
J03285216+3122453 has measured $JHK_s$ within 0.19 mag at all bands to
LAL98, not LAL 120. Given this evidence, I have opted to combine LAL
98 and 120 into 2MASS J03285216+3122453.
\item LAL 108 I strongly suspect to be identical with LAL 139, which
is matched to 2MASS J03285521+3125223. LAL 139 itself does not seem to
be offset; this must be the boundary between tiles in the original
observation. The reported photometry for LAL 108 and 139 are nearly
identical (matching at better than 0.08 mag, all bands).
\item I investigated whether it was reasonable to shift LAL 120 to
become identical to LAL 147, because such a shift would be in the same
direction. However, it is right on top of 2MASS J03285216+3122453, so
I left it at that location. Confusingly, LAL 120 has photometry that
is much different at $J$ from LAL 147, but matches to within 0.13 mag
at $K$; LAL 120 and 2MASS J03285216+3122453 have differing $JHK$
magnitudes by $>$4 mag. I have made no net changes as a result of this
consideration.
\item LAL 124 moves from having no match to 2MASS
J03285630+3124432. LAL dos not report $J$ or $H$, but these sources
match each other at $K$ to 0.13 mag.
\item LAL 134 I strongly suspect to be identical to LAL 165, which is
well-matched to 2MASS J03285842+3122567. The two LAL sources have no
$J$ or $H$, but match each other in $K$ to 0.09 mag; they match the
2MASS $K_s$ to within 0.06 mags.
\end{itemize}

Just based on patterns, this seems to be the end of the sources I can
(or need to) match in this region. Beyond the apparent corner of the
tile, LAL 147, 166, 164, and 163 are all well-matched to 2MASS
counterparts. I cannot recover LAL 130; there is nothing at that
location in 2MASS or IRAC, nor at a place offset in the same direction
and size as the offsets above. 

These coordinate uncertainties have some more minor repercussions in
the rest of the catalog -- there are things I strongly suspect are
duplicates, as well as sources without counterparts that I can find.
As for ASR, even given intrinsic variability, the LAL observations are
not so deep as to be likely to reach sources undetected in the 2MASS,
2MASS 6$\times$, IRAC, or even the WISE observations.  If there was no
source in the 2MASS, IRAC, or WISE images at the LAL position or
within a `reasonable' distance, I often dropped the LAL source; see
Table~\ref{tab:lalmods}. 

In the end, I made changes to 37 LAL sources. Sources that are
duplicates are explicitly indicated in the last column in
Table~\ref{tab:crossids} via notation like `==LAL44' for LAL 38.

\clearpage

\subsection{X-ray catalogs}

Using a deep Roentgen Satellite (ROSAT) High Resolution Imager (HRI)
observation, Preibisch (1997) detected 20 X-ray sources, 16 of which
were taken to be likely cluster members. Getman \etal\ (2002) report
on a Chandra observation of this region, detecting 127 sources, 95 of
which were identified with known cluster members. I retained only the
source numbers (not the fluxes) reported in Preibisch (1997). Getman
\etal\ (2002) find matches to all Preibisch sources except sources 1
and 2. I was able to locate a counterpart at multiple other bands for
source 1, and no counterpart at any band for source 2. I suspect that
source 2 from Preibisch (1997) may be spurious (or extragalactic), and
I have removed it from my catalog.

Some of the sources from Getman \etal\ (2002) were specifically tagged
as variable sources (in X-rays). No new YSO candidates were identified
in their catalog, though it is quite likely many of the newly
identified sources are members.  They reported their entire catalog,
which I absorbed into our database for the cross-IDs, though I did
not retain the X-ray fluxes or luminosities in favor of our own
re-reduction of the Chandra data (R14). 

I found two inconsistencies in the Getman \etal\ (2002) crossmatching
between this catalog and LAL and ASR, which I corrected in the
catalog. Many of the cross-matches in Getman \etal\ (2002) are
correct. However, LAL 79 is much closer to ASR 126 than 127, which is
not what is reported in Getman \etal\ (2002). LAL 93 moved as per the
discussion in Section \ref{sec:applal} above; therefore LAL 93 should
not actually be matched to Getman source 15. Getman source 15 should
be matched to 2MASS J03270047+3137259.

Preibisch (2003) reports on an XMM-Newton observation of NGC 1333. It
covers a wider area than the Chandra data from Getman \etal, but most
of the sources identified in this paper are also identified in Getman
\etal. There are 7 sources identified based solely on the XMM data.
Preibisch finds counterparts in the optical or IR for 5 of the
sources. Given the coordinates in Preibisch, 2 of the objects are
within 1 arcsec of objects in our database, and the rest are within 2
arcsecond of objects in our database, including the two claimed not to
have a counterpart. I have made these associations in our catalog.
Names from this study are incorporated into ours, including for the
two objects previously claimed to not have counterparts. 

The XMM-Newton Serendipitous Source Catalogue 3XMM-DR4 was
released in 2013, and consists of source detections over most
of the XMM data taken as of 2012 December, which includes the NGC 1333
observations from Preibisch (2003). I find 46 matches to sources in
this region, only 10 of which do not already have a Chandra detection
in our reprocessed data. Names for these sources are included in the
`other names' field, preceeded by `XMM.'

There are no sources for which I now have an X-ray measurement (from
YSOVAR's reprocessing or from 3XMM-DR4) that did not already appear in
Preibisch (1997), Getman \etal\ (2002), Preibisch (2003), or Winston
\etal\ (2009, 2010), described below. 

\subsection{Rodr\'iguez \etal\ (VLA) Catalog}

Rodr\'iguez \etal\ (1999) report on Very Large Array (VLA)
observations at 6 and 3.6 cm of an $8\arcmin\times8\arcmin$ region
centered on the HH 7-11 region. The nomenclature for this catalog as
established in the paper is not ``RAC'' (as one might assume) but
rather ``VLA.'' Most of the objects were identified with a YSO
counterpart in this paper; some were identified as variable. I
retained nomenclature and positions from this catalog, because they
are still in wide usage today.

Finding counterparts between 3.6 and 6 cm objects and near- to mid-IR
sources can be quite complicated, as there is no assurance that the
emitting source is the same. I have thus retained some VLA sources
without shorter wavelength counterparts, and I have shifted some
sources  to match shorter wavelength counterparts. As mentioned above
in Sec.~\ref{sec:appasr}, VLA 27 has been tied (including by Rebull
\etal\ 2003) to one of the ASR sources. I have now tied VLA 27 to that
source (R15-NGC1333 J032911.24+311831.8=c2d J032911.3+311831), but not
any of the ASR sources. Also, I have tied VLA 43 to ASR 114 (=2MASS
J03290149+3120208 = R15-NGC1333 J032901.53+312020.6) and VLA 42 to
2MASS J03290116+3120244 (R15-NGC1333 J032901.16+312024.4). Based on
long-wavelength information from Herschel (in the context of the
Sadavoy \etal\ (2014) data incorporation below), I have associated VLA
28 with R15-NGC1333 J032912.05+311301.4. 

\subsection{Other NIR and MIR data}

Rebull \etal\ (2003) reported on ground-based MIR data, and has
largely been superceded by Spitzer data. However, the source
cross-identifications and the source multiplicities found there are
useful, and have been retained in the catalog.  For example, SVS 12
may be extended, and SVS 16 and ASR 107 are both multiple sources.

Aspin (2003) reported on NIR spectroscopy, including spectral types
which I have included in our database.  Some types are not
particularly precise (``early K'') but for many stars, this is all
that is available.

Wilking \etal\ (2004) revisited the cluster with $JHK$ over a
$\sim11\arcmin\times11\arcmin$ region to $K\sim16$. They also searched
for brown dwarfs, reporting on spectral types for their 25 candidates.
The catalog abbreviation for this study, as established in their
paper, is ``MBO,'' standing for Mount Bigelow Observatory. In the
published Wilking \etal\ (2004) catalog, there are two MBO 221s. Both
have (separate) IRAC counterparts, only one has a 2MASS counterpart,
and it is faint at that (it comes from the 2MASS 6$\times$ catalog).
In consultation with B.~Wilking (priv.\ comm.), the source on the west
was manually added. MBO 222 is the highest number in the published
catalog, so this source is now assigned to be MBO 223.   Thus, MBO 221
is now matched to R15-NGC1333 J032847.19+311845.9  and MBO 223 is now
matched to R15-NGC1333 J032847.27+312310.0.  For completeness, and for
the benefit of future users of the catalog, I note that there are also
several MBO numbers missing entirely.

\subsection{Cryo-era Spitzer Catalogs}

As noted above, the first Spitzer data for the NGC 1333 region were
taken were taken cooperatively between the GTOs and the c2d Legacy
team. The Legacy team mapped the entire Perseus molecular cloud
complex, including NGC 1333; J{\o}rgensen \etal\ (2006) reported on
the IRAC data, and Rebull \etal\ (2007) reported on the MIPS data.
Notably, the c2d data delivery included ``bandfilled flux densities,''
meaning that if the source was not detected, an aperture was laid down
at the location of the source to obtain an upper limit. I included
flux densities and upper limits from the c2d catalog where
measurements were not already present in our reprocessing of the
cryogenic-era data.  (This was the case where there was a low
signal-to-noise ratio detection, or a limit.) Note that our cryo
reprocessing does not include 70 or 160 \mum\ fluxes; where relevant
(rarely, in this very crowded region), I obtained 70 \mum\ 
measurements from the c2d catalog.  The c2d project also identified
YSO candidates using a multiband color selection; I retained in the
database an indication of this status. 

Gutermuth \etal\ (2008) reported on the Spitzer data specifically for
NGC 1333 (as opposed to the entire Perseus cloud as c2d did). Only the
candidate young stars were reported in Gutermuth \etal\ (2008). Since
these cryogenic observations were reprocessed for YSOVAR using the
same approach as Gutermuth \etal\ (2008), I only retained
identifications of YSO candidates from Gutermuth \etal\ (2008).  

J{\o}rgensen \etal\ (2007) reported on submillimeter data combined
with the Spitzer c2d data. They included a list of embedded YSOs in
Perseus. These cross-IDs are included in `other names' as J07-xx. 

Arnold \etal\ (2012) included data from a Spitzer Infrared
Spectrograph (IRS) survey of objects in NGC 1333. From this article,
I have retained cross-IDs, SED classes (for comparison to ours), and
identifications of YSOs (and non-members).

\subsection{Hatchell \etal\ (2007) catalog}

Hatchell \etal\ (2007) observed Perseus at 450 and 850 \mum. For some
sources, they compiled SEDs and made cross-identifications. I
included the cross-identifications in our catalog, as well as made new
associations between these sources and and the short-wavelength
counterparts. 

Source number 49 is listed as a match to an IRAS source, as well as
the 2MASS source J03283609+3113346. Looking at the 2MASS and WISE
images, this is an extended source, and that 2MASS source is one of
three near the core of it. The 2MASS point source closest to the core
of the extended source is J03283681+3113326; the 2MASS extended source
catalog source to which it should be matched is J03283630+3113329.

Source number 70 does not have a short-wavelength counterpart listed
in Hatchell \etal. There is nothing obvious there in the 2MASS images,
and only faintly is there a source in the WISE images. There is a
source from Spitzer/IRAC at J032914.96+312031.7, which is 4.4$\arcsec$
away from the position given for source 70, within the range of
positional uncertainties given by the Hatchell \etal\ cross-references
to 2MASS. Hatchell \etal\ place this source as a SED Class 0.  I have
very few points between 2 and 25 \mum\ delineating J032914.96+312031.7
-- I only have four IRAC bands, and it is falling, not rising. I place
it in SED Class II.   The Hatchell \etal\ source 70 may not be a good
match to the source at R15-NGC1333 J032914.96+312031.7, but it is the
best thing I can match to it at this time. A note about this
uncertainty is included in the last column of
Table~\ref{tab:crossids}. 

\subsection{Brown Dwarf Searches}

There are several studies seeking brown dwarfs in this region. Oasa
\etal\ (2008) is one of these brown dwarf searches, using $JHK_s$ over
a $\sim$5$\arcmin\times$5$\arcmin$ region with a spatial resolution of
$\sim$1.5$\arcsec$. Their YSO candidate identifications were retained
in my catalog. Matches were found within 2$\arcsec$ for 85\% of their
sources. The rest of the sources were retained in the catalog with the
positions from Oasa \etal, except for one, which I now describe.

I find that one of the sources reported upon in Oasa \etal\ (2008) is
subject to source confusion issues. It is in the region associated
with IRAS-2b. The source in question was assigned a position of
03:28:57.09 +31:14:21.4 by Oasa \etal, who reported no other sources
within 40$\arcsec$ of this location. In the IRAC images, two sources
can be seen, J032857.37+311415.7 and J032857.20+311419.1, both of
which are quite far away from the source from Oasa \etal\
(7.4$\arcsec$ and 3.4$\arcsec$, respectively).  They have very
different flux ratios -- the more northern one dominates at $\sim$3
\mum\ (and is closer to the original Oasa position), but by $\sim$8
\mum, the southern one dominates.  By inspection of the available
images (2MASS, IRAC, and WISE), I conclude that this Oasa \etal\
source should not be retained as a separate source, but instead be
tied to one of the other two seen in the IR. Based on proximity, I
have tentatively tied this source to R15-NGC1333 J032857.20+311419.1.
However, Oasa \etal\ report that this source is a Class III, and not a
YSO. This is not consistent with the rest of the information I have on
J032857.20+311419.1, but there aren't any closer sources bright in the
NIR. This is the best association I can make. A note about this
uncertainty is included in the last column of
Table~\ref{tab:crossids}. 

\begin{figure}[ht]
\epsscale{0.8}
\plotone{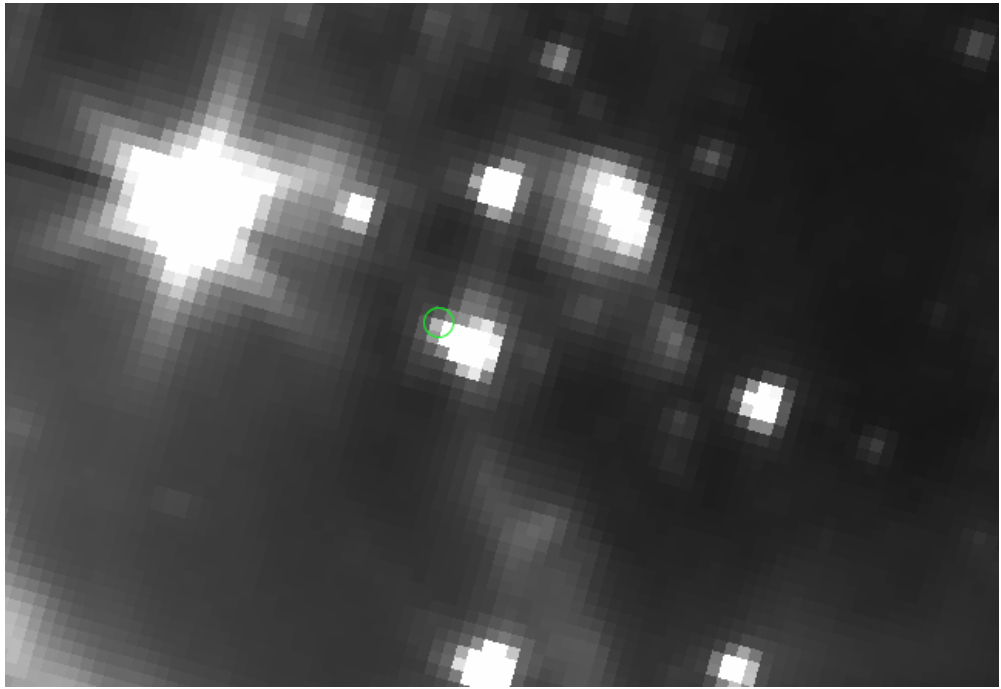}
\caption{The region of sky in 3.6 \mum\ immediately around 
J032857.41+311536.9=Greissl S1-6, circled.  North is up. The
distinctive triangularly shaped IRAC point spread function of the
point sources can be seen in several point sources in this field of
view, but one has an `appendage' extending to the left; the
`appendage' is matched to Greissl S1-6.  }
\label{fig:greissls1-6}
\end{figure}

Greissl \etal\ (2007) used the Hubble Space Telescope (HST) NICMOS
instrument to search for brown dwarfs in NGC 1333. They reported $JH$
photometry and spectral types from low-resolution grism spectroscopy
for their objects of interest, both of which I included in the
database.  Greissl \etal\ source S1-6 is very close to other sources
in the region; Figure~\ref{fig:greissls1-6} shows the region in 3.6
\mum. There is an `appendage' off of a brighter source; that
`appendage' is what I matched to source S1-6=J032857.41+311536.9. The
net SED for this source is largely optical/NIR data, from Greissl
\etal; it is hard to apportion IR data correctly for this source from
telescopes other than HST. I considered that this source may be
incorrectly matched to an IR source, and that the optical points may
correspond to a different source. The nearest one I can identify is
the brighter source in Fig.~\ref{fig:greissls1-6}
(J032857.15+311534.6), which is $\sim$4$\arcsec$ away; this is
significantly  larger than the IR positional uncertainties, or that of
Greissl \etal, so this is unlikely. Moreover, since they were working
with HST data, it is likely that their astrometry is correct. I have
left it in the catalog as-is. There are notes about this in the last
column of Table~\ref{tab:crossids} for S1-6=J032857.41+311536.9.

Scholz \etal\ (2009), Scholz \etal\ (2012a), and Scholz \etal\ (2012b)
are all part of the SONYC (Substellar Objects in Nearby Young
Clusters) survey searching for brown dwarfs.  These investigators
obtained very deep imaging in four bands ($i^{\prime},z^{\prime},J,K$)
but only report photometry for their brown dwarfs, candidates, and
rejects, which I incorporate into the catalog.   I also included the
spectral types and membership as reported in the SONYC papers, both
for stellar and substellar objects. Scholz \etal\ (2012b) report a new
very low mass member, which becomes S-45. However, it is within 0.2
arcsec of Sp105, a source they list as a previously confirmed member
in Scholz \etal\ (2012a). I take these two to be referencing the same
object. They have essentially identical NIR magnitudes and spectral
types in the papers.

Itoh \etal\ (2010) is another search for brown dwarfs;
spectral types and cross-identifications from this paper were
incorporated into the database.

\subsection{Winston catalog}

Winston \etal\ (2009, 2010) report on a combined Spitzer and Chandra
investigation of NGC 1333. Follow-up spectra were also included. A
full X-ray catalog (of every source detected) was included in Winston
\etal\ (2010). Spectral types and X-ray identifications were
incorporated from this paper into our database, as well as whether
Winston \etal\ identified the object as a YSO. The X-ray measurements
from Winston \etal\ were superceded by our own re-reduction (R14). 

The source numbers presented in Winston \etal\ (2009) refer to
source numbers tied to coordinates (RA/Dec) in Winston \etal\ (2010),
but there is some potential confusion as to whether the numbers are
sequential IR or X-ray numbers, where the former come from Gutermuth
\etal\ (2008), and the latter are assigned in the Winston work. The
numbers in Winston \etal\ (2009) are not specified as being IR or
X-ray numbers. Winston \etal\ (2010) reports both X-ray and IR
numbers, but not all of them appear in Gutermuth \etal\ (2008). Via
personal communication with E.~Winston, I have verified that the
numbers in Winston \etal\ (2009) are meant to be IR numbers. The IR
numbers in Winston \etal\ (2009) and (2010) that are greater than 137
do not appear in Gutermuth \etal\ (2008). These are
ones that Winston \etal\ assigned (not Gutermuth \etal) and are
defined via postions (RA/Dec) in Winston \etal\ (2009).  Thus, while
Gutermuth \etal\ (2008) only assigns numbers to YSO candidates
reported there, there can be a ``Gutermuth08 number'' assigned by
Winston \etal\ (2009, 2010). In other words, Winston \etal\ (2009)
reports those as YSO candidates selected from the IR, but Gutermuth
\etal\ (2008) does not report those objects as YSO candidates.  All of
this has been resolved in my catalog.

\subsection{Sadavoy et al. Herschel identifications}

\begin{figure}[ht]
\epsscale{0.8}
\plotone{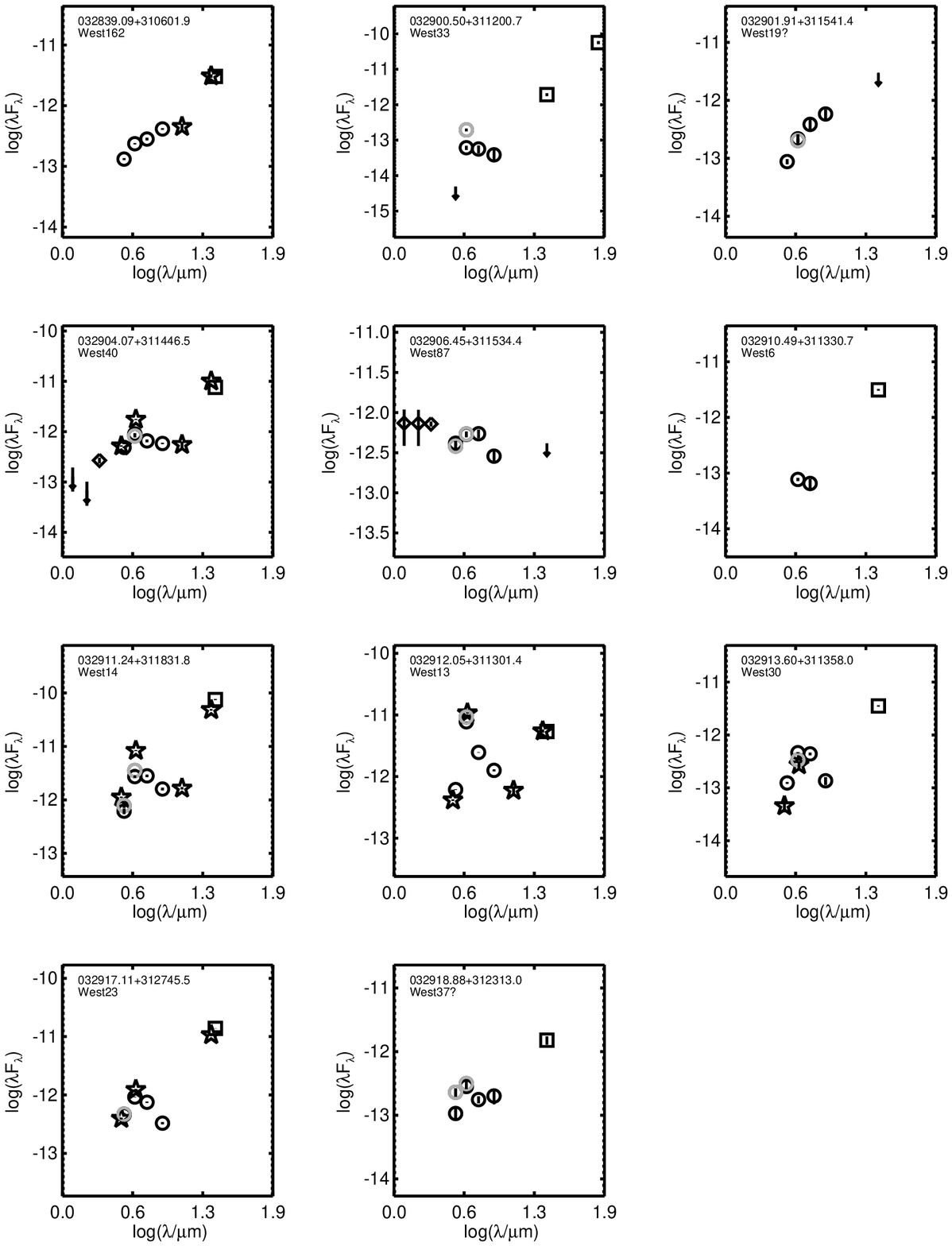}
\caption{SEDs for the sources matched to the sources from Sadavoy
\etal\ (2014). Plots are log $\lambda
F_{\lambda}$ in cgs units (erg s$^{-1}$ cm$^{-2}$) against log
$\lambda$ in microns.  Diamonds= 2MASS, circles=IRAC (with grey
circles being mean from YSOVAR campaign), squares=MIPS, stars=WISE,
and downward arrows=upper limits. Small vertical lines in each point
are the error bars. R15-NGC1333 catalog numbers appear in the plots.
See the text for details of source matching,
missing bands, etc. }
\label{fig:sadavoyseds}
\end{figure}

Sadavoy \etal\ (2014) report on candidate Class 0 objects in NGC 1333
using Herschel data. Because the shortest Herschel wavelength, 70
\mum, overlaps with the longest Spitzer wavelength I have here (also
70 \mum), I wished to link the Herschel observations with the rest of
my catalog. However, this is another example of the difficulty in
finding clear correspondences between sources at shorter and longer
wavelengths. In many cases, the corresponding source at 3.6 and/or 4.6
\mum\ is blended with nebular emission and may not represent flux
solely from the source seen at long wavelengths.  Some of the sources
in Sadavoy  \etal\ in this region are noted as not having Spitzer
counterparts, but I have been able to find counterparts for most of
them.  All of these cross-identifications appear as part of the `other
names' in Table~\ref{tab:crossids}; where the association is
uncertain, a question mark is added to the cross-identification. No
flux densities are reported in Sadavoy \etal\ (2014), so these points
cannot be added to tables or SEDs. My assembled SEDs for these objects
appear in Figure~\ref{fig:sadavoyseds}. Notes on specific sources
follow. 

Counterparts to Sadavoy \etal\ sources West162 and West33 are likely
to have contamination from extended emission at 3.6 and 4.5 \mum.
Moreover, for West33, the short-wavelength counterpart has a falling
SED, but by 24 \mum, is is very bright and rising fast. The source
detected at 3-8 \mum\ may not be the same object seen at $>$24 \mum\
in this case.

West19 is very complicated because it is in a very bright and confused
region. There does seem to be an object (J032901.91+311541.4) at IRAC
wavelengths close to that location; at 7.5$\arcsec$, it is among the
largest positional offsets for all of the matches to the Sadavoy
sources, but consistent with the postional uncertainties.  At 24 \mum,
the object at this location is very unfortunately placed in the MIPS
image so as to be largely obscured by the Airy ring of a nearby
extremely bright source. (Note that I have not tied West19 to the
very bright 24 \mum\ source, since it is too far away!)  This
association is the best I can do at this time, despite the relatively
large positional offset.

West40 and West87 are both in the same neighborhood as West19, though
they are both farther from the very bright sources and extended
emission that plague West19. By contrast, West40's match to
J032904.07+311446.5 seemes relatively straightforward, though the
$J$-[24] SED is somewhat unusually shaped. West87's match at
J032906.45+311534.4 does not have a [24] measurement because of the
halo around the bright 24 \mum\ source. It has another strange SED,
and the object responsible for the emission at the shortest
wavelengths may very well not be the same as for the $>$20 \mum\
wavelengths. Because it is missing SED points longer than 20 \mum, and
because of the 2-8 \mum\ slope, I place this object in SED Class II.

For West6, there is a clear correspondence of the 24 \mum\ source (at
J032910.49+311330.7) and the Herschel source. However, no clear 8
\mum\ source can be seen on the IRAC image (and no 12 \mum\ source can
be seen on the WISE image). There is a faint source at 4.5 \mum, and
perhaps something on the 5.8 \mum\ image, but nothing clear on the 3.6
\mum\ image. It is not apparent if the source at 4.5 (and 5.8) \mum\ is
the same as the 24 \mum\ source and West6, but all measurements have
been retained in this associated source.

West14 is in another very complicated region. There are several
sources in close proximity as well as extended nebular emission. I
have tied it to R15-NGC1333 J032911.24+311831.8 on the basis of image
morphology at Spitzer bands, but note that there could be
contributions from other adjacent sources at nearly all bands. (For
example, WISE may blend the two closest sources.)

West13 is another object that is difficult to match to short
wavelength sources. The 70 \mum\ source seen by Herschel has a clear
counterpart in the 24 \mum\ MIPS images. However, there may be more
than one source in the IRAC images at this location. Based on the
images, I associate West13 with R15-NGC1333 J032912.05+311301.4 as the most
likely match. Based on this association, I move VLA 28 from its
nominal published location to coincide with this source as well. There
is a nearby source seen in 2MASS 6$\times$, but only at $K_s$. This
source, however, is too far away to be associated with
R15-NGC1333 J032912.05+311301.4, so I do not associate them.

West30 and West23 are both additional cases of a clear match among the
70, 24, 8, and 5.8 \mum\ images, but the measurements at 3.6 and 4.5
\mum\ may be affected by extended emission.

West37 is another very complicated, bright region. There is clearly an
object in the 3.6, 4.5, and 5.8 \mum\ images, but if I had only those
short wavelengths, I might have called it a dust clump based on image
morphology. There is a source faintly seen in the 24 \mum\ image, but
is on a nearby source's Airy ring. I have provisionally tied West37 to
R15-NGC1333 J032918.88+312313.0.

\subsection{Foster APOGEE data}

Foster \etal\ (2015) report on the velocity dispersion of young stars
in NGC~1333 based on near-IR spectra obtained with the Apache Point
Observatory Galactic Evolution Experiment (APOGEE; Zasowsky \etal\
2013). Target selection for that project included considerations based
on the variability of objects in the YSOVAR data, so many of the
objects in which we are interested in the context of YSOVAR also have
measurements in Foster \etal. Through spectral fitting, Foster \etal\
derive several parameters for stars with sufficient signal-to-noise
ratios in their APOGEE spectra, including  \teff, log $g$, and
\vsini.  I included in our database cross-identifications,
identifications of non-members, \teff, and \vsini, as well as other
information from that paper.

\subsection{YSOVAR data}

Because I had originally undertaken this project in the context of
analysis of our YSOVAR data, I have taken the liberty in this paper
of including the means from those light curves (at 3.6 and 4.5 \mum)
in our database. As mentioned above, the NGC 1333 YSOVAR data will be
described in detail in Rebull \etal\ (2015 in prep).

\subsection{Two Very Bright, Confused Sources}

In one of the brightest regions of NGC 1333, there are two sources
very close to each other, at or near the spatial resolution of many of
the surveys here. Early on, this pair of sources was identified as SVS
12=IRAS 6. However, in 2MASS, IRAC, and even in MIPS images, two
sources can be distinguished by eye if the stretch is severe enough.
Both of these objects have slightly different SEDs, certainly in part
due to saturation and flux apportionment issues. The one tagged ASR
114=LAL 181=MBO 19=Gutermuth 27=c2d J032901.6+312021=R15-NGC1333
J032901.53+312020.6 made it onto most published lists as a YSO or
candidate. For reasons probably due to saturation and/or spatial
resolution and/or flux apportionment, the second did not; it is c2d
J032901.2+312025=R15-NGC1333 J032901.16+312024.4. In the c2d catalog,
one of these source gets all the MIPS 70 \mum\ flux density, though
the other source probably should be allocated some of it. Many bands
are listed as limits for these sources in various catalogs; they are
probably meant as lower, not upper, limits, though they are often
tagged upper limits. I have declared both of these sources to be
literature-identified YSOs, and added a note about the 70 \mum\ flux
apportionment to the last column in Table~\ref{tab:crossids}. 

\subsection{Jet Drivers}

I consulted two papers to identify the sources most commonly thought
of as driving outflows and/or jets in NGC 1333. Davis \etal\ (2008)
used 2.122 \mum\ imaging to identify outflows; they identified 11
sources as having outflows of any sort, and their IDs were tied back
to those from J{\o}rgensen \etal\ (2007).  Plunkett \etal\ (2013)
reported on CARMA observations of outflows, identifying seven YSOs as
the drivers of the outflows. Three sources are identified by both
papers, resulting in 12 sources identified as driving jets or
outflows.  These sources, as well as some of their other names,
are included in Table~\ref{tab:jetdrivers}.

\begin{deluxetable}{cll}
\tablecaption{List of Jet/Outflow Driving Sources \label{tab:jetdrivers}}
\tablewidth{0pt}
\tablehead{\colhead{Our ID (R15-NGC1333 J)} & \colhead{Reference(s)} 
& \colhead{Some of the Other IDs}  }
\startdata
032837.06+311331.0 & Davis \etal\ (2008) & IRAS1, IRAS 03255+3103, J07-11\\
032845.30+310542.0 & Davis \etal\ (2008) & IRAS 03256+3055, J07-14\\
032855.53+311436.3 & Plunkett \etal\ (2013), Davis \etal\ (2008) & IRAS 2A, SK8, J07-15\\
032857.37+311415.7 & Davis \etal\ (2008) & IRAS 2a, J07-16\\
032900.50+311200.7 & Plunkett \etal\ (2013), Davis \etal\ (2008) & IRAS 4B1, SK1, Sadavoy2014-West33, J07-18\\
032903.39+311602.0 & Davis \etal\ (2008) & J07-20 \\
032903.75+311603.9 & Plunkett \etal\ (2013) & SVS13, IRAS3, IRAS 03259+3105,\\
032904.07+311446.5 & Plunkett \etal\ (2013) & IRAS5, Sadavoy2014-West40, SK14, J07-21\\
032910.49+311330.7 & Plunkett \etal\ (2013), Davis \etal\ (2008) & IRAS4A, SK4, Sadavoy2014-West6,J07-22\\
032910.70+311820.9 & Davis \etal\ (2008) & J07-23\\
032911.24+311831.8 & Davis \etal\ (2008) & Sadavoy2014-West14,J07-24 \\
032912.05+311301.4 & Plunkett \etal\ (2013) & IRAS4B, SK3, Sadavoy2014-West13, J07-25\\
\enddata
\end{deluxetable}

\clearpage

\subsection{Studies That Are Not Included}

There are several individual famous objects of particular interest in
this region, some with many papers entirely of their own (e.g., the
components that make up IRAS-2, IRAS-4, \ldots). I did not include
additional data from these many individual projects in our catalog,
but instead assume that users can match by cross-id to any few sources
of interest (though I encourage image inspection via the IRSA tool
FinderChart if nothing else). I also did not place an emphasis on
matching to many surveys at wavelengths longer than about 50 \mum, 
because (a) there are relatively few true point sources at the longer
bands, and (b) it is often hard to ensure a good match between the
point sources from shorter wavelengths and those at the longer
wavelengths. Some long-wavelength matches have been made in three
cases -- the VLA catalog is the source of some nomenclature still in
wide use, and both the Hatchell \etal\ (2007) and Sadavoy \etal\
(2014) papers made specific efforts to find matches at the shorter
wavelengths. A comprehensive set of matches to all of the sub-mm and
mm sources in the literature is beyond the scope of this paper.

\section{Obvious Gaps for Future Work}
\label{sec:future}

About half of the objects in this region that are identified as YSO
candidates lack spectroscopy. This could be used not just for spectral
classification, but for confirmation that they are young stars, or
even just confirmation that they are not background galaxies (this is
a very real concern; see, e.g., Rebull \etal\ 2010).  Most of the
spectral types found in the literature come from searches for brown
dwarfs, so the set of objects for which there is spectral types is
incomplete and highly skewed to late M and later.  There are about 150
objects (out of $\sim$300 YSO candidates) that have some sort of
estimate of spectral type in the literature, even a coarse one (e.g.,
``$<$M0" or ``early K''). The \teff\ values from Foster \etal\ (2015)
can be used to constrain the spectral type, even though the \teff\
values are much more uncertain for the hotter stars. However, there
are only about a 20 objects for which there is a \teff\ estimate but
no published spectral type of any kind. Spectroscopy would help limit
the non-member contamination and improve the inventory of members (and
therefore knowledge of the mass function) in NGC 1333.

Multi-band broadband optical data could be very helpful over this
entire region for delineating the Wien side of the SED. Even just $r$
and $i$ (or similar) would help show if the SEDs are, in fact, turning
over for the YSO candidates. Moreover, short wavelengths such as $u$-
and $g$-band data would be useful for constraining mass accretion
rates in these young stars, but such observations would be quite
challenging due to the high extinction towards most sources. High
spatial resolution optical observations have been shown (e.g., Rebull
\etal\ 2010) to be critical for distinguishing background star forming
galaxies from nearby YSOs, since both kinds of objects overlap in IR
color space with IR colors indicating star formation.

X-ray data can be very helpful for identifying young stars without 
disks.  The existing Chandra and XMM data are limited in area, focused
just  on the heart of NGC 1333.  X-ray data over a larger area could
help identify additional less-embedded cluster members. 

Some of the most famous YSOs in this region are those originally
identified by the Infrared Array Satellite (IRAS) in 1983. However,
most of these very bright sources have broken into pieces every time
astronomers have looked with higher spatial resolution. Some sources
are still clearly multiple and still have flux apportionment problems.
Higher spatial resolution MIR and FIR observations will aid in
correctly attributing source flux to the correct source. In some
cases, the surface brightness is so high that higher spatial
resolution observations could be very difficult.

\section{Placement into SED Classes}
\label{sec:sed}

After assembling all of the literature data, including the mean [3.6]
and [4.5] measurement from our YSOVAR campaign, I constructed an SED
for each object using all available data. Some objects have
well-defined SEDs, with data from optical to 8, 24, or even 70 \mum;
others have far less-well-defined SEDs, for example, with only one
point that is the mean of one channel's light curve from our YSOVAR
campaign.

As discussed in R14, I defined an internally consistent placement of
the YSOVAR objects into SED classes as follows. In the spirit of
Wilking \etal\ (2001), I define the near- to mid-IR (2 to 24 \mum)
slope of the SED, $\alpha = d \log \lambda F_{\lambda}/d \log 
\lambda$,  where  $\alpha > 0.3$ for a Class I, 0.3 to $-$0.3 for a
flat-spectrum  source, $-$0.3 to $-$1.6 for a Class II, and $<-$1.6
for a Class III.  For each of the YSOs and candidate YSOs in the
sample, I performed a simple ordinary least squares linear fit to {\bf
all available photometry} (just detections, not including upper or
lower limits) {\bf as observed between 2 and 24 $\mu$m, inclusive}. 
Note that formal errors on the IR points are generally so small as to
not affect the fitted SED slope. Note also that the fit is performed
on the observed SED, e.g., no reddening corrections are applied to the
observed photometry before fitting. Classification via this method is
provided specifically to enable comparison to other YSOVAR papers via
internally consistent means; see discussion in R14. I can only perform
this calculation for those objects with points at more than one
wavelength in their SED between 2 and 24 \mum. Therefore, objects for
which I have, e.g., one 3.6 \mum\ point from the cryogenic era and one
3.6 \mum\ YSOVAR point (the mean YSOVAR measurement at this band)
cannot have a fitted SED slope.  

The SED slopes and classes I calculate appear in
Table~\ref{tab:bigdata}. Slopes can be calculated for any of the
objects with at least two distinct wavelengths in the SED between 2
and 24 \mum, but it is only meaningful if the source is a YSO, so I
only include slopes and a class for objects tagged as YSOs by any of
the references I included in this paper.  Of the $\sim$300 objects in
the catalog that are tagged as YSOs, where I can calculate the slope
between 2 and 24 \mum, I find 55 Class Is, 38 Flat class, 103 Class
IIs, and 81 Class IIIs. There are 25 with insufficient points in the
SED, such that I cannot assign a class. 

\begin{figure}[ht]
\epsscale{0.8}
\plotone{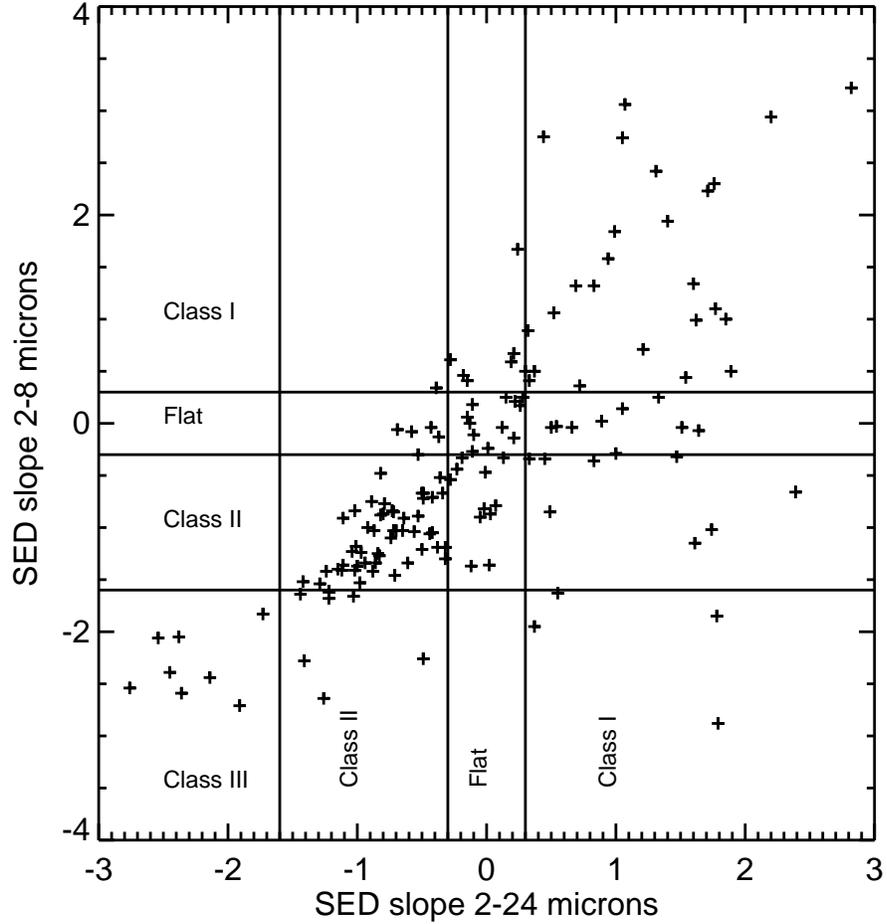}
\caption{A comparison of the SED slopes calculated from the observed
SED, just for those objects with 24 \mum\ detections and identified
somewhere in the literature as YSOs.  Since the 24 \mum\ detections
are strongly biased towards cluster members, most of the points here
are YSOs (or candidates). The vertical and horizontal lines indicate
the divisions between SED classes as defined in Sec.~\ref{sec:sed}.
For 65\% of the entire set of objects, the resulting SED class is the
same. The objects that do not match typically have a borderline slope.
As expected, there is a bias such that inclusion of the 24 \mum\ point
frequently pushes an object to more embedded SED classes. }
\label{fig:compareslopes}
\end{figure}

In R14, we asserted that the classes are generally well matched
whether or not the 24 \mum\ point is included. To understand the
influence of the 24 \mum\ point, in the cases where there is a 24
\mum\ point, I can compare the 2-24 \mum\ and 2-8 \mum\ SED slopes. 
Figure~\ref{fig:compareslopes} compares the fitted slope for the 2-24
\mum\ and 2-8 \mum\ approaches, just for those sources detected at 24
\mum, and also identified as candidate YSOs. (The 24 \mum\ detections
are strongly biased towards cluster members, so this figure includes
most of the 24 \mum\ detections.) The vertical and horizontal lines
indicate the divisions between SED classes as defined above. For
$\sim$65\% of this subset of objects, the resulting SED class is the
same. (Table~\ref{tab:sedclasses} has the total numbers.) The objects
that do not match typically have a borderline slope. As expected,
there is a bias such that inclusion of the 24 \mum\ point frequently
pushes an object to more embedded SED classes. There are a few sources
that are approximately photospheric until a sharp rise at 24 \mum;
those are difficult to classify correctly using our approach.   Of the
$\sim$300 things identified as YSOs (or candidates), $\sim$150 are
detected at 24 \mum, and $\sim$100 have the same class even including
the 24 \mum\ point.

In R14, we also asserted that our classes are generally well-matched
to the G09 classes.  Arnold \etal\ (2012) also report classes for YSO
candidates. Table~\ref{tab:sedclasses} has the numbers of objects in
our catalog for which my classes match (or do not match) the classes
obtained from these other approaches. The first thing to notice is
that both G09 and Arnold \etal\ (2012) have different class bins than
I have defined -- G09 has no flat class, and a ``II/III'' class, and
Arnold \etal\ have transition disk (TD) and pre-transition disk (PTD)
classes. Despite this, the majority of the sources have the same class
regardless of approach. 

For the most embedded sources, there is no provision in our scheme for
identifying Class 0s, and I do not use points at wavelengths longer
than 24 \mum\ to determine classes. I would expect, though, that
those objects that others have identified as Class 0s would work out
to be Class Is in my scheme. There are 10 sources with SED
classifications from the submm in Hatchell \etal\ (2007) for which I
have counterparts -- in Hatchell \etal, eight of them are Class 0 and
two are Class I, and by my classification, nine are Class I and one
is Class II. There are eleven Class 0 sources in the region identified
in Sadavoy \etal\ (2014) from Herschel data -- out of those 11, all
but one are Class I, but that last one is Class II.  Both of those
discrepant sources (J032914.96+312031.7 for the Hatchell source and
J032906.45+311534.4 for the Sadavoy source) have been discussed above.
The matches may not be a good match to the source of the long
wavelength flux. At any rate, in most cases, my method at least
recovers Class I status for these very embedded objects.

Ideally, one would have a fully-populated SED, as well as a spectral
type, and thus be able to attempt SED modelling to determine the proper
classes for these objects, but this is not possible at this time.

\begin{deluxetable}{rcccc}
\tablecaption{Comparison of Classes \label{tab:sedclasses}}
\tablewidth{0pt}
\tablehead{\colhead{Other class} & \colhead{Our Class I} 
& \colhead{Our Flat Class} & \colhead{Our Class II} 
& \colhead{Our Class III} }
\startdata
Literature tagged YSOs: 2-8 \mum\ Class I    & 25 &  7 & 1 & 0 \\
Literature tagged YSOs: 2-8 \mum\ Flat Class &  9 & 12 & 5 & 0 \\
Literature tagged YSOs: 2-8 \mum\ Class II   &  8 &  11 & 49 & 0\\
Literature tagged YSOs: 2-8 \mum\ Class III  &  4 &  0 & 8 & 8 \\
\hline
Any matches: G09 Class I,I* & 29 & 9 & 0 & 0\\
Any matches: G09 Class II,II* & 5 & 19 & 67 & 3 \\
Any matches: G09 Class II/III & 0 & 0 & 4 & 0 \\
\hline
Any matches: Arnold12 class I  & 12 & 2 & 0 & 0\\
Any matches: Arnold12 class FS & 2 & 7 & 2 & 0\\
Any matches: Arnold12 class II & 0 & 1 & 33 & 0\\
Any matches: Arnold12 class III& 0 & 0 & 0 & 1\\
Any matches: Arnold12 class PTD& 0 & 1 & 3 & 0\\
Any matches: Arnold12 class TD & 0 & 2 & 3 & 0\\
\enddata
\end{deluxetable}

\include{bigclasscatalogabbrev}

\section{Summary}
\label{sec:concl}

I have presented a catalog of apparent point source objects towards
NGC 1333, within $52\arcdeg<$RA$<52.5\arcdeg$ and
$31\arcdeg<$Dec$<31.6\arcdeg$.  I have attempted to resolve a wide
variety of confusion in the literature, ranging from mismatched
sources to missing or duplicate IDs. I incorporated data from optical
to radio wavelengths, but focused most of my effort on $J$ (1.25
\mum) to 24 \mum. Cross-identifications include those from more than
25 papers and catalogs from 1994-2015. 

I have also identified obvious holes in the accumulated literature,
the most obvious one being spectroscopic confirmation of the many YSO
candidates presented in the literature, along with spectral types. 

While I have done the best that I can, the reliability of this
catalog is likely lower for the longer wavelengths, and for the most
bright and/or confused regions in the heart of NGC 1333. Image
inspection proved invaluable in making many of these associations.

I have compared a few different methodologies for classifying objects
by the SED shape and/or IR colors. While the methods agree in most
cases, they can fail in identifying the most embedded sources, and in
sources that are on the borderline between SED class slope
definitions, and can more weakly depend on whether or not there is a
detection at $\sim$20-25 \mum\ to anchor the SED slope between 2 and 24
\mum.

We will use this catalog as the basis for our upcoming work using
YSOVAR data in NGC 1333.

\acknowledgments

Special thanks to all of the authors of past studies who patiently
answered my questions, dug up old data and notes, and helped me sort
out which object was which. Thanks to David Shupe and the
NASA-Herschel Science Center helpdesk for a quick processing of PACS
and SPIRE images to help resolve source matching issues. Thanks to
Jesus Hernandez for pointing out the problem between WISE All-Sky and
AllWISE for the sources in IRAS 7 and to Chris Gelino for helping
resolve it. Thanks to John Stauffer, Moritz G\"unther, and Lynne
Hillenbrand for comments on the manuscript.

This research has made use of the NASA/IPAC Infrared Science Archive
(IRSA), which is operated by the Jet Propulsion Laboratory, California
Institute of Technology, under contract with the National Aeronautics
and Space Administration.   This work is based in part on observations
made with the Spitzer Space Telescope, which is operated by the Jet
Propulsion Laboratory, California Institute of Technology under a
contract with NASA. Support for this work was provided by NASA through
an award issued by JPL/Caltech.  This research has made use of NASA's
Astrophysics Data System (ADS) Abstract Service, and of the SIMBAD
database, operated at CDS, Strasbourg, France.  This research has made
use of data products from the Two Micron All-Sky Survey (2MASS), which
is a joint project of the University of Massachusetts and the Infrared
Processing and Analysis Center, funded by the National Aeronautics and
Space Administration and the National Science Foundation. The 2MASS
data are served by the NASA/IPAC Infrared Science Archive, which is
operated by the Jet Propulsion Laboratory, California Institute of
Technology, under contract with the National Aeronautics and Space
Administration. This publication makes use of data products from the
Wide-field Infrared Survey Explorer, which is a joint project of the
University of California, Los Angeles, and the Jet Propulsion
Laboratory/California Institute of Technology, funded by the National
Aeronautics and Space Administration.

\end{document}